\documentclass[twocolumn]{aastex63}

\received{}
\revised{}
\accepted{}
\submitjournal{ApJ}

\shorttitle{Shock-cloud interaction in the RCW 86 southwestern rim}
\shortauthors{Suzuki et al.}
\usepackage{threeparttable}
\usepackage{multirow}
\usepackage{color}
\usepackage{amsmath}
\usepackage{breakcites}

\begin{document}

\title{Particle acceleration controlled by ambient density in the southwestern rim of RCW~86}

\correspondingauthor{H. Suzuki}
\email{hiromasa050701@gmail.com}

\author[0000-0002-8152-6172]{Hiromasa Suzuki}
\affiliation{Department of Physics, Faculty of Science and Engineering, Konan University, 8-9-1 Okamoto, Higashinada, Kobe, Hyogo 658-8501, Japan}
\nocollaboration{1}

\author{Satoru Katsuda}
\affiliation{Graduate School of Science and Engineering, Saitama University, 255 Shimo-Ohkubo, Sakura, Saitama 338-8570, Japan}
\nocollaboration{1}

\author[0000-0002-4383-0368]{Takaaki Tanaka}
\affiliation{Department of Physics, Faculty of Science and Engineering, Konan University, 8-9-1 Okamoto, Higashinada, Kobe, Hyogo 658-8501, Japan}
\nocollaboration{1}

\author{Nobuaki Sasaki}
\affiliation{Graduate School of Science and Engineering, Saitama University, 255 Shimo-Ohkubo, Sakura, Saitama 338-8570, Japan}
\nocollaboration{1}

\author[0000-0002-7935-8771]{Tsuyoshi Inoue}
\affiliation{Department of Physics, Faculty of Science and Engineering, Konan University, 8-9-1 Okamoto, Higashinada, Kobe, Hyogo 658-8501, Japan}
\nocollaboration{1}

\author[0000-0002-5456-4771]{Federico Fraschetti}
\affiliation{Center for Astrophysics $|$ Harvard \& Smithsonian, Cambridge, MA, 02138, USA}
\affiliation{Department of Planetary Sciences, Lunar \& Planetary Laboratory, University of Arizona, Tucson, AZ 85721, USA}
\nocollaboration{1}



\begin{abstract}
Particle acceleration physics at supernova remnant (SNR) shocks is one of the most intriguing problems in astrophysics.
SNR RCW~86 provides a suitable environment for understanding the particle acceleration physics because one can extract the information of both accelerated particles and acceleration environment at the same regions through the bright X-ray emission.
In this work, we study X-ray proper motions and spectral properties of the southwestern region of RCW~86.
The proper motion velocities are found to be $\sim 300$--2000~km~s$^{-1}$ at a distance of 2.8~kpc.
We find two inward-moving filaments, which are more likely reflected shocks rather than reverse shocks.
Based on the X-ray spectroscopy, we evaluate thermal parameters such as the ambient density and temperature, and non-thermal parameters such as the power-law flux and index.
From the flux decrease in time of several non-thermal filaments, we estimate the magnetic field amplitudes to be $\sim 30$--100~$\mu$G.
Gathering the physical parameters, we then investigate parameter correlations.
We find that the synchrotron emission from thermal-dominated filaments is correlated with the ambient density $n_{\rm e}$ as $\text{(power-law flux)} \propto n_{\rm e}^{1.0 \pm 0.2}$ and $\text{(power-law index)} \propto n_{\rm e}^{0.38 \pm 0.10}$, not or only weakly with the shock velocity and shock obliquity.
As an interpretation, we propose a shock-cloud interaction scenario, where locally enhanced magnetic turbulence levels have a great influence on local acceleration conditions.

\end{abstract}



\section{Introduction} \label{sec-intro}
Particle acceleration physics at supernova remnant (SNR) shocks is one of the principal problems in astrophysics as a promising mechanism to produce cosmic rays below the ``knee'' energy ($\approx 3 \times 10^{15}$~eV).
X-ray and gamma-ray studies of SNRs have revealed several aspects of the particle acceleration physics in SNRs.
\cite{volk05} and \cite{vink06b} suggested that magnetic field amplification is very effective in young ($< 2$~kyr) SNRs.
%In SN~1006, the X-ray roll-off energy was found to depend clearly on the shock obliquity, indicating high acceleration efficiencies at parallel shocks \citep{miceli09}.
High turbulence levels of magnetic fields in some parts of young SNRs were found with the highest level close to the Bohm limit at $\sim 2$~kyr \citep{tsuji21}.
\cite{reynolds21} studied the energy amount of accelerated electrons and magnetic field in young SNRs based on several physical parameters such as radio luminosity, plasma density, and shock velocity, finding large variations in them among objects controlled by unknown factors.
\cite{suzuki22a} also found that the maximum energies of accelerated protons differ by more than one order of magnitude among objects at similar ages.

As the essential part of the mechanism to accelerate particles up to the knee energy, the enhancement of magnetic-field strength and turbulence has been attracting particular interest (e.g., \citealt{bell04, bamba05, vink06, amato06, uchiyama07}).
As a possible cause of such a field enhancement, shock-cloud interactions are thought to be important \citep{giacalone07, inoue12, fraschetti13}.
The magnetic turbulence is expected to be amplified around dense clumps and enhanced synchrotron X-rays have been in fact observed (e.g., \citealt{sano13, sano15, sano17}).
%In order to reach an understanding of the microphysics of particle acceleration, both thermal and non-thermal information of the acceleration sites are essential as well as shock velocities.

SNR RCW~86 provides a suitable environment for understanding the particle acceleration microphysics.
One can extract the information of both accelerated particles and acceleration environment at the same regions because the bright X-ray emission exhibits both thermal and non-thermal components.
RCW~86 is believed to be the remnant of the oldest historical supernova of A.D. 185 \citep{stephenson02, green03}.
RCW~86 is located at ($l, b$) = (315.4, $-$2.5) and has a radio shell with a radius of $\sim 21'$, which is almost completely surrounded by Balmer-dominated filaments \citep{smith97}.
The distance is estimated to be 2.8~kpc \citep{rosado96}.
RCW~86 is thought to have evolved in a low-density cavity region and is currently interacting with dense materials \citep{williams11}.
The highly irregular morphology of the SNR shell indicates that RCW~86 is currently expanding in a very inhomogeneous ambient medium.
Such inhomogeneity yields a broad range of shock velocities and magnetic turbulence around the shocks that will affect the efficiency of particle acceleration.
The north-eastern (NE) corner is thought to be expanding in a rather low-density medium with large velocities \citep{yamaguchi16} and emits hard non-thermal X-rays \citep{bamba00b, vink06}.
On the other hand, the radio-brightest south-western (SW) corner is likely interacting with a dense cloud that modifies the shock structure, thereby reducing or enhancing the non-thermal X-ray emission \citep{rho02, sano17, sano19b}.

In this work, we focus on the SW region. We investigate the shock velocities, spectral features around shocks, and particle acceleration environments.
The observation logs and data reduction process are described in Section~\ref{sec-obs}.
Our analysis procedure and results are presented in Section~\ref{sec-analysis}.
The shock structures and acceleration environments are discussed in Section~\ref{sec-discussion}.

\section{Observations and Data Reduction} \label{sec-obs}
We use all the four existing Chandra observations of the RCW~86 SW region listed in Table~\ref{tab-log}.
The baseline for the proper motion study is $\approx 12$ year, which consists of the first-epoch observation in 2001 (OBSID 1993) and the second-epoch ones in 2013 (OBSIDs 13748, 15610, and 15611).
The total exposure time is 177~ksec.

The RCW~86 SW region is observed with the Advanced CCD Imaging Spectrometer (ACIS; \citealt{garmire97}) S2, S3, and I3 in 2001 and with S2, S3, and I2 in 2013.
All the data were taken in the FAINT mode. We process the raw data following the standard data reduction method ({\tt chandra\_repro}).
We used CIAO (v4.11; \citealt{fruscione06}) and calibration database 4.8.3 for the data reduction.

\begin{table*}[htb!]
\centering
\caption{Chandra observation logs of the RCW~86 SW region
\label{tab-log}}
\begin{tabular}{l l l l l l l}
\hline\hline
OBSID & R.A. (2000.) & Decl. (2000.) & Roll angle & Date & Exposure (ksec) & PI \\ \hline
1993 & 220.19279 & $-$62.66287 & 80\fdg2 & 2001 Feb 01 & 92 & S. Reynolds \\
13748 & 220.11392 & $-$62.71971 & 70\fdg7 & 2013 Feb 14 & 36 & S. Katsuda \\
15610 & 220.11389 & $-$62.71975 & 70\fdg7 & 2013 Feb 17 & 26 & S. Katsuda \\ 
15611 & 220.11392 & $-$62.71971 & 70\fdg7 & 2013 Feb 12 & 23 & S. Katsuda \\ \hline
\end{tabular}
\end{table*}

\section{Analysis and Results} \label{sec-analysis}
We perform proper motion study, spectroscopy, and filament-width measurement for the RCW~86 SW region.
The procedures and results are presented in this section.
In our analysis, we use HEAsoft (v6.20; \citealt{heasarc14}), XSPEC (v12.9.1; \citealt{arnaud96}), and AtomDB 3.0.9.
Throughout the paper, uncertainties in the text, figures, and tables indicate $1\sigma$ confidence intervals.

\begin{figure*}[htb!]
\centering
\includegraphics[width=15cm]{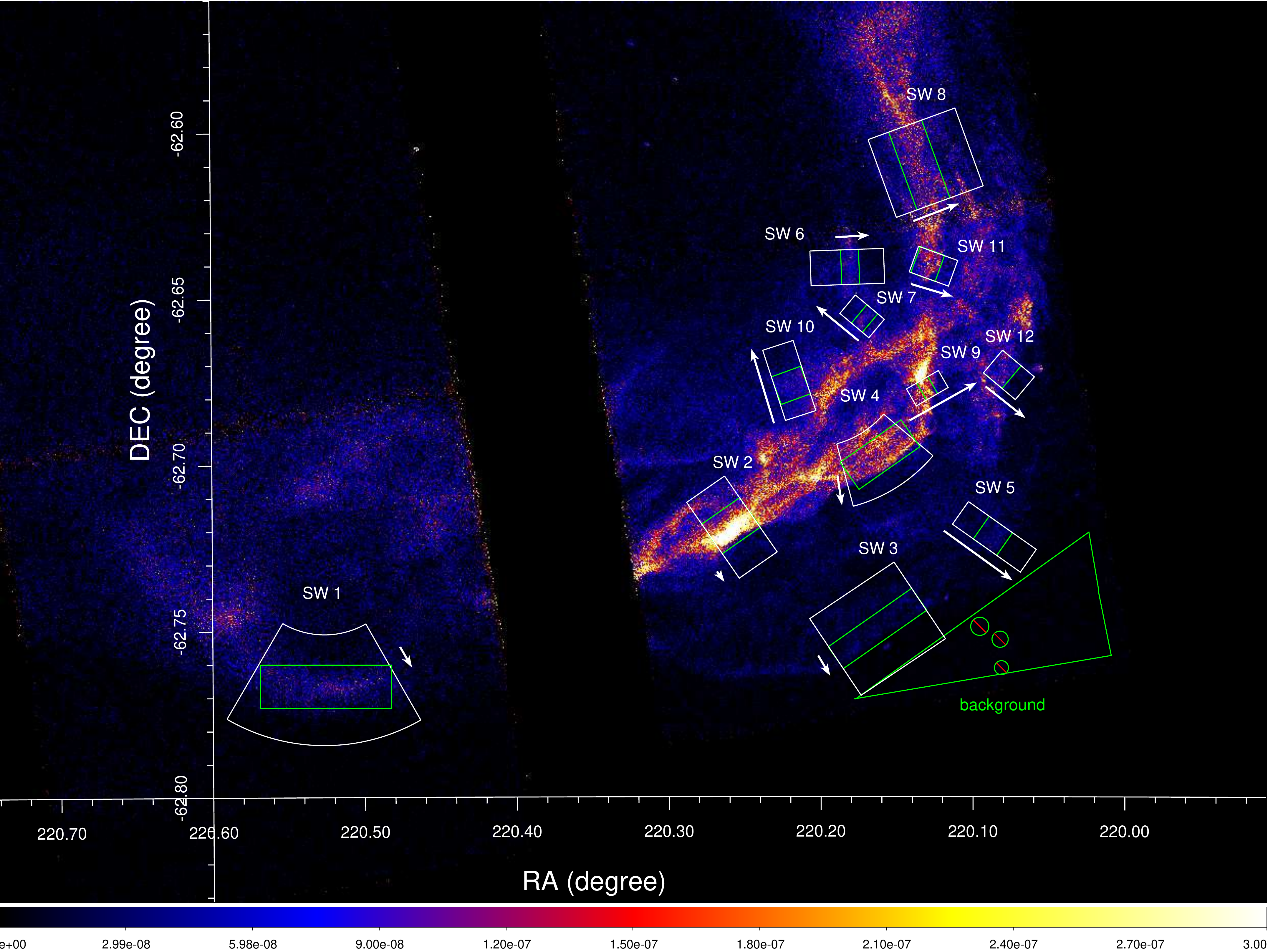}
\caption{Exposure-corrected 0.5--7.0 keV image obtained with Chandra in 2001. The white and green lines indicate the analysis regions for proper motion and spectral studies, respectively.
The arrows beside the regions indicate the moving directions {and velocities (proportional to the arrow lengths)} of the filaments as described in Section~\ref{sec-prop}.
Their directions also correspond to the positive directions of the positions in Figures~\ref{fig-profile1} and \ref{fig-profile2} in the Appendix~\ref{sec-prop2}.
\label{fig-image}}
\end{figure*}

\begin{figure}[htb!]
\centering
\includegraphics[width=8cm]{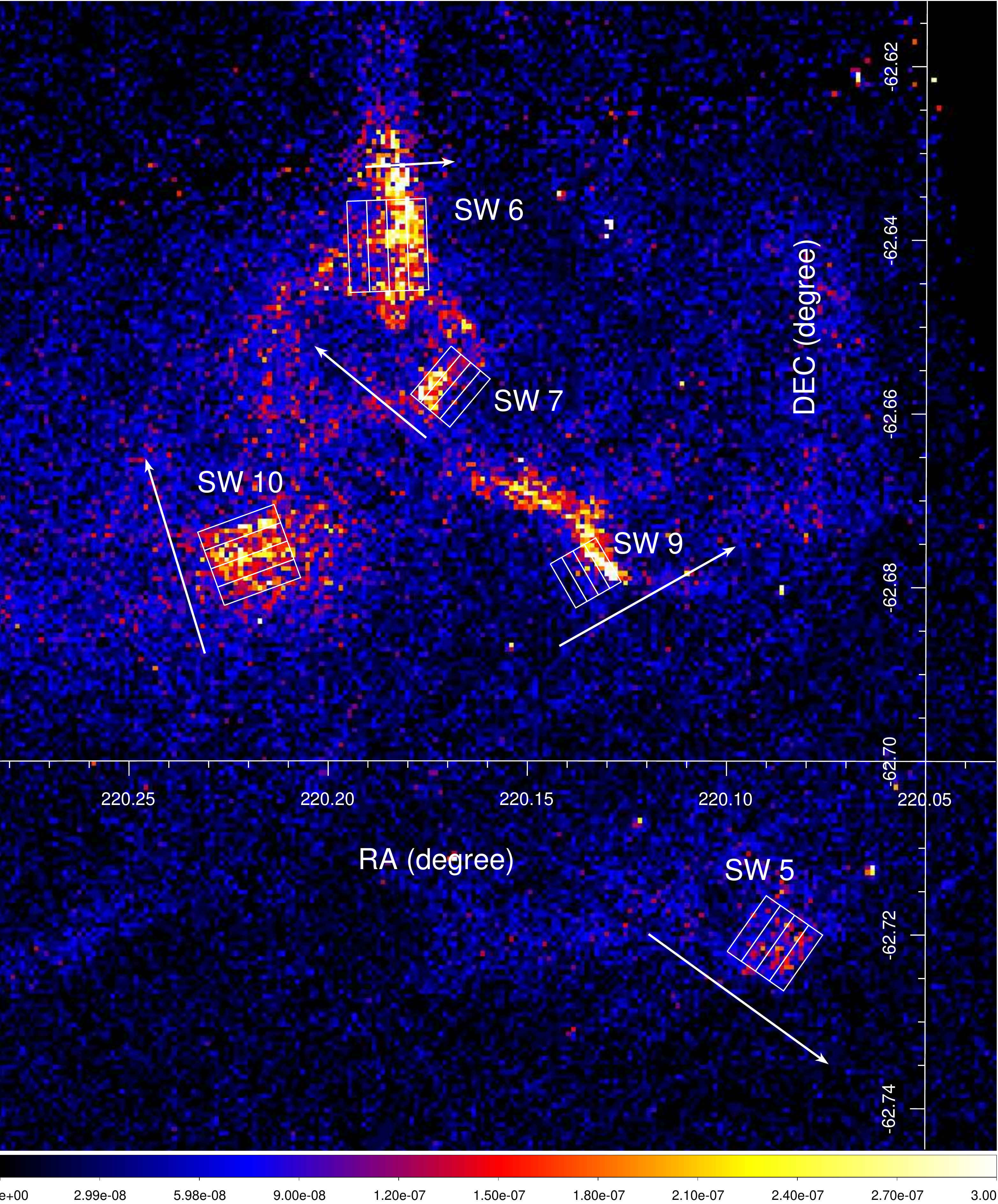}
\caption{Exposure-corrected 2.0--7.0 keV image in 2001.
The white rectangular regions are used for the non-thermal spectral variation study.
The arrows beside the regions indicate the moving directions {and velocities (proportional to the arrow lengths)} of the filaments as described in Section~\ref{sec-prop}.
The arrows also correspond to the positive directions of the angular positions in Figure~\ref{fig-hardening}.
\label{fig-image2}}
\end{figure}

\begin{figure}[htb!]
\centering
\includegraphics[width=8cm]{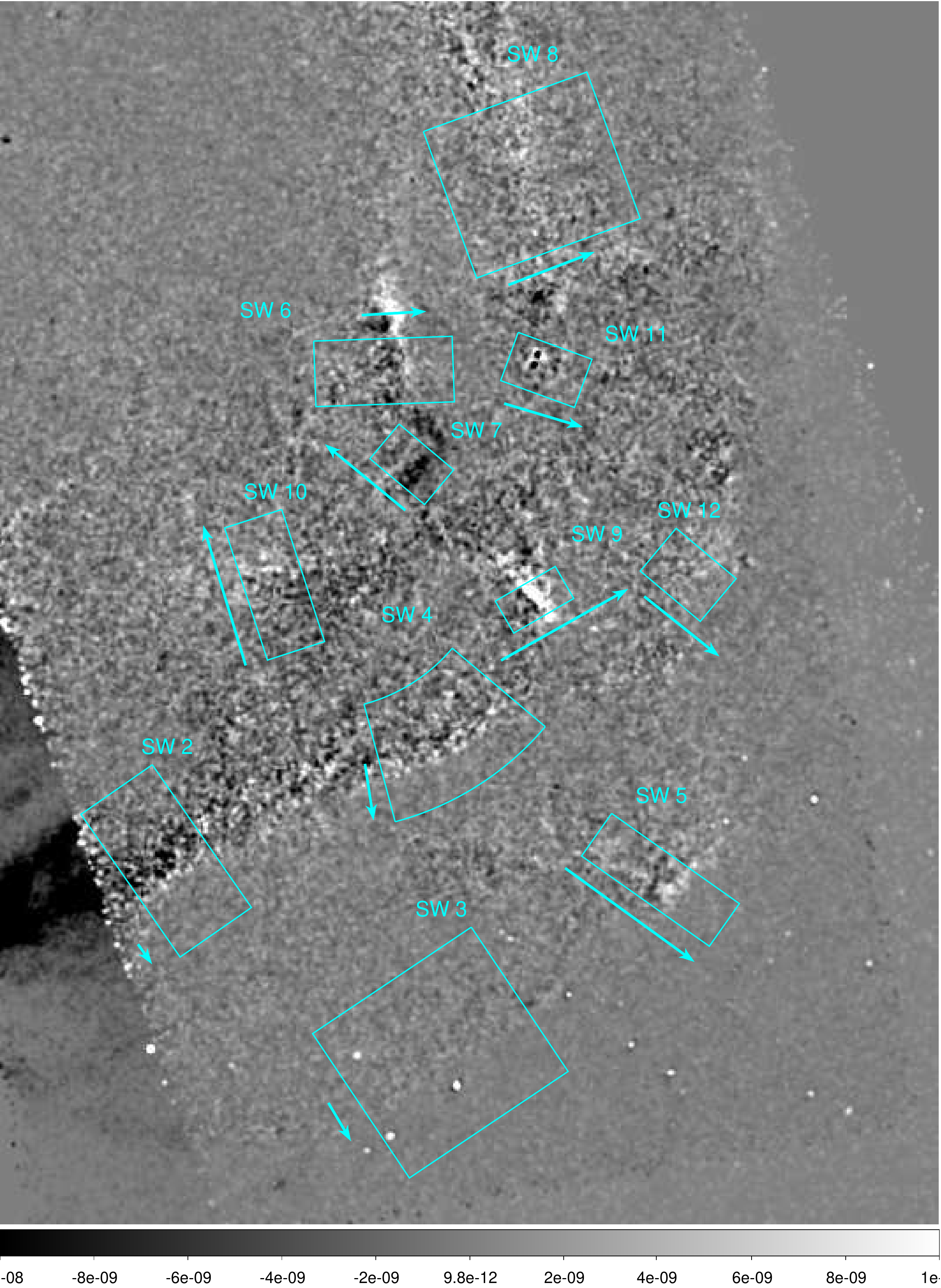}
\caption{{The image difference of the exposure-corrected 1.0--5.0 keV images between 2001 and 2013.
The image is smoothed with a Gaussian kernel of $\sigma = 0\farcs8$.
The white and black pixels show increased and decreased fluxes from 2001 to 2013, respectively.
The cyan regions and arrows indicate the analysis regions and proper motion directions and velocities of the filaments.
Note that a triangular region at the bottom-left corner is covered only by the observation in 2001.}
\label{fig-subt}}
\end{figure}

{The wide-band (0.5--7.0~keV) and hard band (2.0--7.0 keV) images extracted from the observation in 2001 are presented in Figures~\ref{fig-image} and \ref{fig-image2}, respectively}, with an indication of the analysis regions.
{Note that only SW5, SW6, SW7, SW9, and SW10 are prominent in the hard X-ray image.
We also make an image showing the difference between the exposure-corrected 1.0--5.0~keV fluxes in 2001 and 2013 (Figure~\ref{fig-subt}).
For some filaments such as SW6, SW7, and SW9, their motions are visible in this image.}
{These 12 analysis regions are selected to enclose all the bright and sharp filament structures seen in the field of view. The flux-profile extraction directions are defined by eye to approximately match the directions perpendicular to the filament structures.}

\subsection{Proper motion measurements}

\subsubsection{Aspect Correction}
\begin{figure}[htb!]
\centering
\includegraphics[width=8cm]{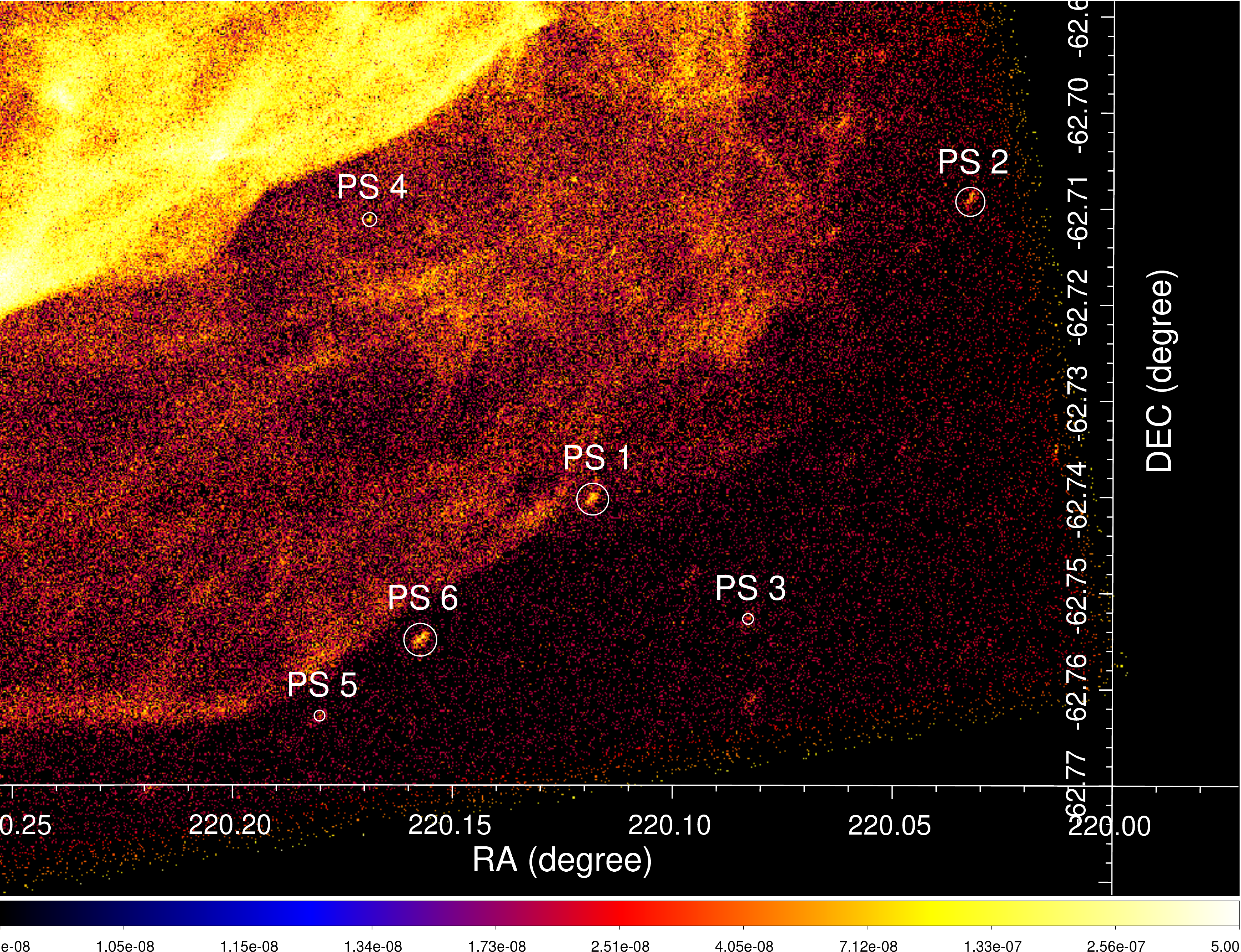}
\caption{Exposure-corrected 0.5--7.0 keV image in 2001 with the six point-like source locations used for aspect correction indicated with white circles. 
\label{fig-ps}}
\end{figure}

In order to perform the proper motion study with the best position accuracy available, we apply the aspect correction to the three observations in 2013.
The six point-like sources indicated in Figure~\ref{fig-ps} are selected and used for the aspect correction because they are bright and close to the nominal points in all the observations.
Before the aspect correction, the positions of these sources differ among the observations typically by $\sim 0\farcs3$.
We first run the {\tt wavdetect} tool for each observation to determine the central positions of these sources, which are summarized in Table~\ref{tab-ps}.
Then we run the {\tt wcs\_match} tool to find the best-fit transformation matrices to correct the coordinate systems of the three observations in 2013 to match those of the observation in 2001\footnote{All the six point sources are used in all the calculations without being excluded in the {\tt wcs\_match} process.}.
The transformation matrix only uses two-dimensional translation without rotation and scaling {because the number of available point-like sources is relatively small}.
Finally, the {\tt wcs\_update} tool applies these corrections to the observations in 2013.

After the correction, for each of the three observations in 2013, the resultant correction accuracy is evaluated based on the standard deviation of the position offsets of the six sources with respect to those in 2001.
The correction accuracies are obtained as $0\farcs18$, $0\farcs24$, and $0\farcs26$, for OBSIDs 13748, 15610, and 15611, respectively.
We merge the three observations in 2013 for use in the following analysis.
The aspect correction accuracy for these merged data is estimated by taking an exposure-weighted average of those for the three observations, to be $0\farcs22$, which is converted to $0\farcs018$ yr$^{-1}$.
This is considered as the systematic uncertainty associated with the position accuracy.

\begin{table*}[htb!]
\fontsize{8}{9}\selectfont
\centering
\caption{Positions of the six point-like sources
\label{tab-ps}}
\begin{tabular}{l l l r r r r r r }
\hline\hline
 & \multicolumn{2}{c}{Position: 1993} & \multicolumn{2}{c}{Diff.: 1993--15610\tablenotemark{a}} & \multicolumn{2}{c}{Diff.: 1993--13748\tablenotemark{a}} & \multicolumn{2}{c}{Diff.: 1993--15611\tablenotemark{a}} \\ 
Source & R.A. (2000.) & Decl. (2000.) & R.A. ($''$) & Decl. ($''$) & R.A. ($''$) & Decl. ($''$) & R.A. ($''$) & Decl. ($''$) \\ \hline
PS1 & 220.1181474 & $-$62.7402549 & $-$0.17 (0.33) & $-$0.04 (0.17) & $-$0.57 (0.29) & 0.35 (0.16) & 0.67 (0.43) & 0.71 (0.17) \\
PS2 & 220.0325486 & $-$62.7092583 & 0.29 (0.32) & $-$0.45 (0.10) & 0.29 (0.35) & 0.03 (0.10) & 0.09 (0.33) & 0.26 (0.10) \\
PS3 & 220.0828321 & $-$62.7526891 &  $-$0.13 (0.26) & $-$0.31 (0.17) & $-$0.57 (0.22) & 0.22 (0.18) & $-$0.39 (0.47) & 0.46 (0.12) \\
PS4 & 220.1688151 & $-$62.7111788 &  $-$0.26 (0.21) & $-$0.08 (0.12) & $-$0.40 (0.21) & 0.45 (0.12) & $-$0.20 (0.21) & 0.64 (0.18) \\
PS5 & 220.1801661 & $-$62.7627948 &  0.64 (0.57) & $-$0.71 (0.26) & 0.00 (0.36) & $-$0.24 (0.17) & 0.13 (0.50) & $-$0.61 (0.24) \\
PS6 & 220.1572650 & $-$62.7548930 &  $-$1.15 (0.46) & $-$0.44 (0.15) & $-$0.84 (0.32) & 0.12 (0.16) & $-$0.62 (0.36) & $-$0.14 (0.21) \\
\hline
\end{tabular}

\tablenotetext{a}{Position differences between two observations in the 0.5--7.0 keV energy range.}

\end{table*}

\subsubsection{Proper motions}\label{sec-prop}
\begin{figure}[htb!]
\centering
\includegraphics[width=8cm]{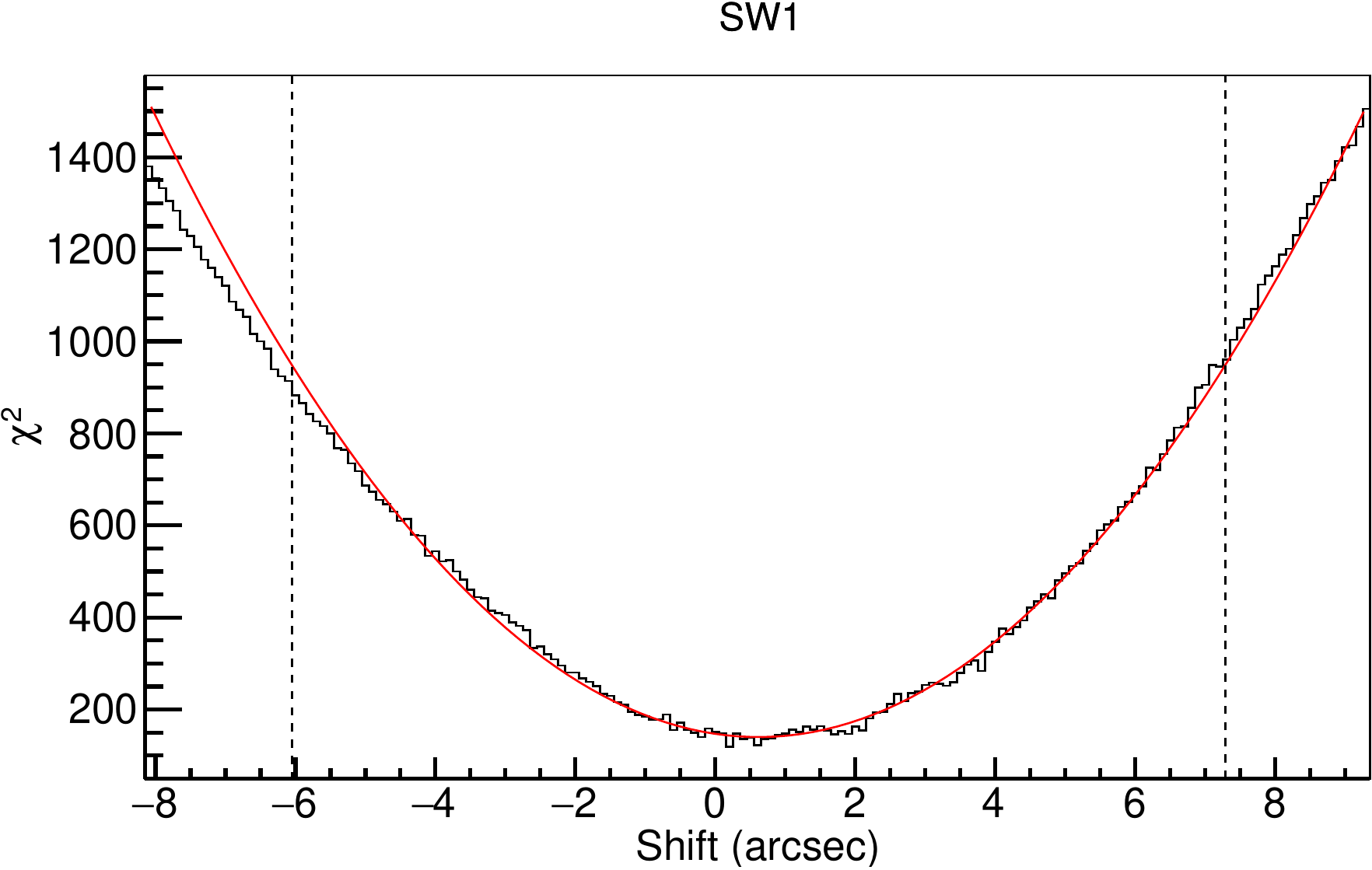}
\caption{Example of the $\chi^2$ test for proper motion measurement for the SW1 region.
The $\chi^2$ values (black solid line) and best-fit parabola model (red line) are shown.
The black dotted lines indicate the angular range for the parabola fitting.
\label{fig-chi2}}
\end{figure}

Proper motion velocities of the filaments indicated in Figure~\ref{fig-image} are measured as follows.
Flux profiles are extracted from the vignetting-corrected images in 2001 and 2013, which are presented in Figures~\ref{fig-profile1} and \ref{fig-profile2}.
To evaluate their proper motion velocities, a $\chi^2$-test is used as below.
We artificially shift the profile in 2013 by $\Delta x$ and calculate $\chi^2 (\Delta x)$, which is defined as 
%\begin{linenomath}
\begin{equation}
\chi^2 (\Delta x) = \sum_{i} \frac{(f_{i} - g (\Delta x)_{i})^2}{\Delta f_{i}^2 + \Delta g (\Delta x)_{i}^2},
\end{equation}
%\end{linenomath}
where $f_i$ and $\Delta f_i$ indicate the flux and error of the bin number $i$ in 2001, and $g_i$ and $\Delta g_i$ indicate those of the shifted profile in 2013.
This calculation is repeated with various values of $\Delta x$ to plot $\chi^2$ as a function of $\Delta x$.
The minimum $\chi^2$ value ($\chi^2_{\rm min}$) and corresponding profile shift ($\Delta x_{\rm min}$) are determined by fitting the $\chi^2$--$\Delta x$ plot with a parabola function.
An example of the $\chi^2$--$\Delta x$ plot and parabola fitting is presented in Figure~\ref{fig-chi2}.
The best-fit $\Delta x_{\rm min}$ is converted to the proper motion velocity.
The profile shift is not limited to an integer multiple of the bin width.
We re-bin the shifted profile $g (\Delta x)$ with the same bin arrangement as $f$ with an assumption of a uniform probability distribution inside each bin.
Then, the profile-shift ranges which give $\chi^2 (\Delta x) = \chi^2_{\rm min} +1$ are calculated from the best-fit parabola functions.
These ranges are considered to be 1$\sigma$ confidence ranges of the proper motion velocities.

\begin{figure}[htb!]
\centering
\includegraphics[width=8cm]{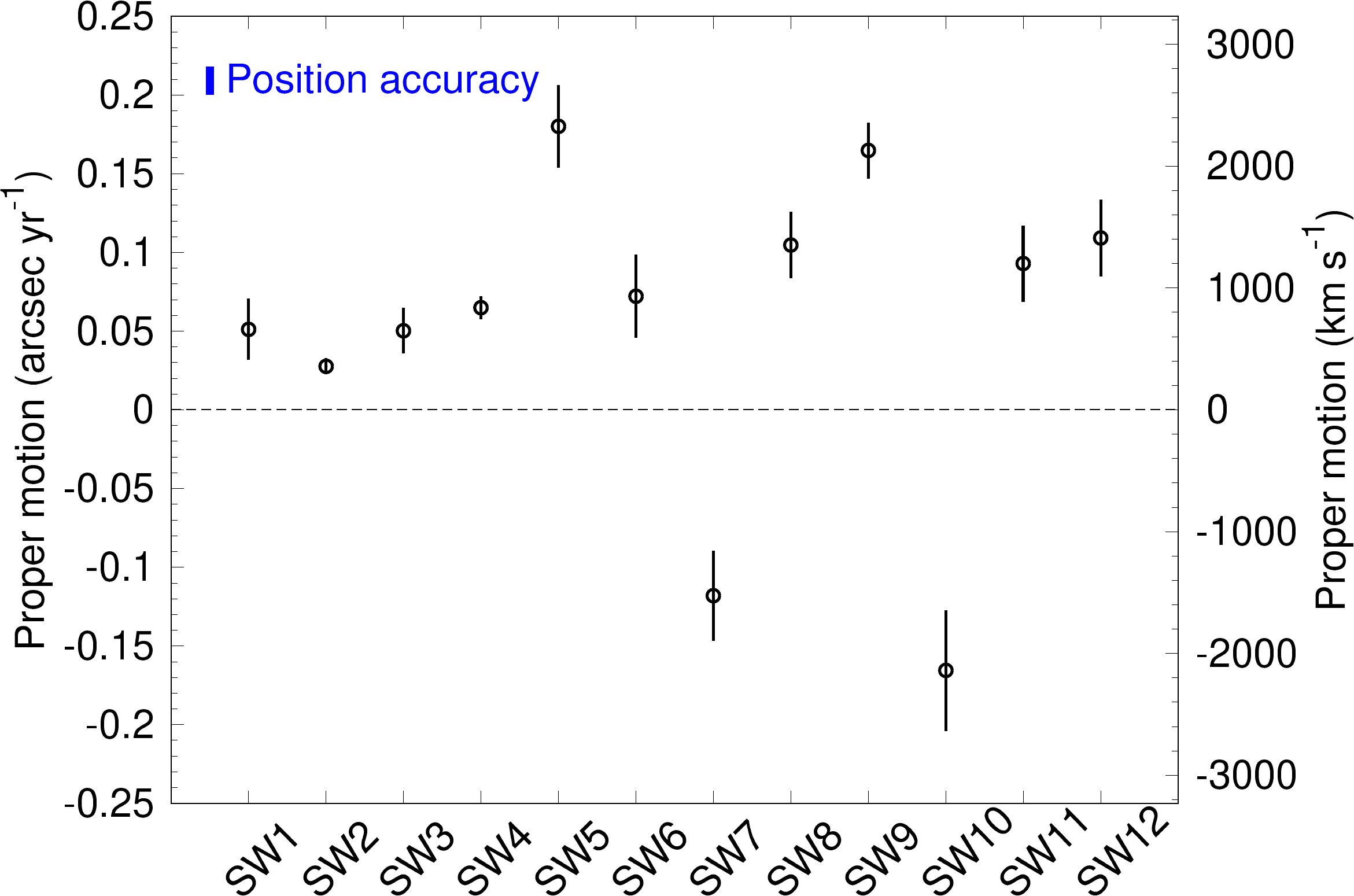}
\caption{Proper motion estimates for individual filaments in units of arcsec~yr$^{-1}$ (on the left) and km~s$^{-1}$ (on the right).
A distance of 2.8~kpc is assumed.
\label{fig-prop}}
\end{figure}

The resultant proper motion estimates are summarized in Table~\ref{tab-prop}, and shown in Figures~\ref{fig-profile1} and \ref{fig-profile2}.
Note that the filaments SW7 and SW10 are moving inward, toward the SNR center.
These inward movements are confirmed by spectral softening toward presumable downstream in SW7 and SW10 as described in Section~\ref{sec-hardening}.
We also note that the projection effect, i.e., the difference between the measured radial velocities and actual three-dimensional velocities, will be small ($\lesssim 15\%$) considering the positions of the filaments with respect to the apparent SNR radius.\footnote{We assume the SNR radius of $\approx 22.6'$ and radial distances of the filaments from the explosion center of $\approx 19.5\text{--}22.6'$.}

\begin{table}[htb!]
%\fontsize{8}{9}\selectfont
\centering
\caption{Proper motion velocities
\label{tab-prop}}
\begin{tabular}{l r r }
\hline\hline
Name & Velocity ($''$~yr$^{-1}$)\tablenotemark{a} & Velocity (km~s$^{-1}$)\tablenotemark{a} \\
\hline
SW1 & $ 0.051 \pm 0.020 $ & $ 660 \pm 250 $ \\
SW2 & $ 0.028 \pm 0.005 $ & $ 360 \pm 70 $ \\
SW3 & $ 0.050 \pm 0.015 $ & $ 650 \pm 190 $ \\
SW4 & $ 0.065 \pm 0.007 $ & $ 840 \pm 90 $ \\
SW5 & $ 0.180 \pm 0.026 $ & $ 2330 \pm 340 $ \\
SW6 & $ 0.072 \pm 0.026 $ & $ 930 \pm 340 $ \\
SW7 & $ -0.118 \pm 0.029 $ & $ -1530 \pm 370 $ \\
SW8 & $ 0.105 \pm 0.021 $ & $ 1350 \pm 270 $ \\
SW9 & $ 0.165 \pm 0.018 $ & $ 2130 \pm 230 $ \\
SW10 & $ -0.166 \pm 0.038 $ & $ -2140 \pm 500 $ \\
SW11 & $ 0.093 \pm 0.024 $ & $ 1200 \pm 310 $ \\
SW12 & $ 0.109 \pm 0.025 $ & $ 1410 \pm 320 $ \\
\hline
\end{tabular}
\tablenotetext{a}{Minus velocities indicate movements toward the SNR center.}
\end{table}

\subsection{Filament Widths and Their Energy Dependence}

\begin{figure*}[htb!]
\centering
\includegraphics[width=14cm]{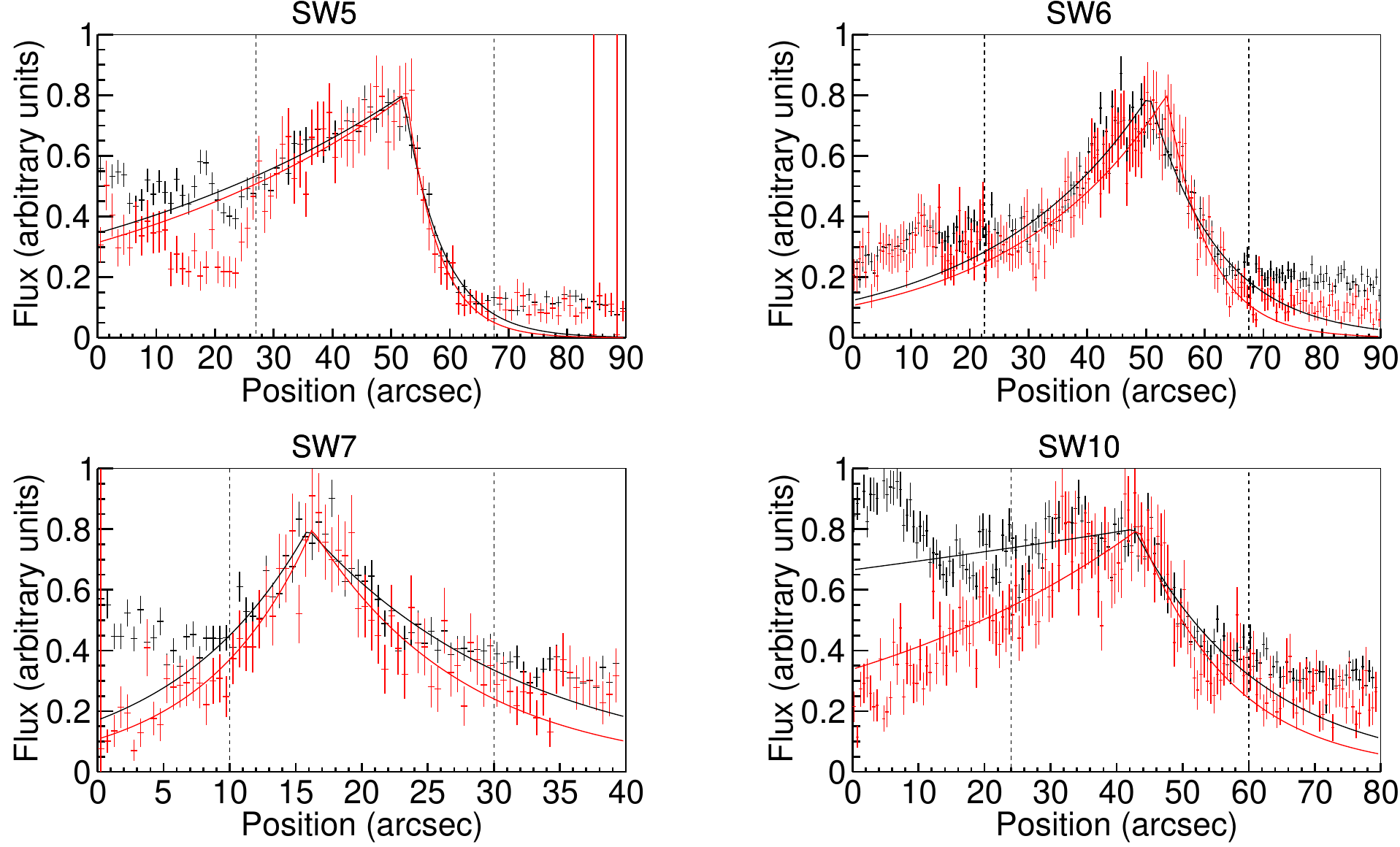}
\caption{Flux profiles of the non-thermal filaments.
The extraction energy ranges are 0.5--2.0~keV (black) and 2.0--7.0~keV (red).
Increasing positions correspond to the directions of the arrows shown in Figure~\ref{fig-image}.
{The displayed flux ranges are different for different panels.}
The best-fit models for the two profiles are overplotted with black and red solid lines, respectively.
Radial ranges used for the fitting are indicated with black dotted lines.
\label{fig-profile-width}}
\end{figure*}

\begin{figure*}[htb!]
\centering
\includegraphics[width=16cm]{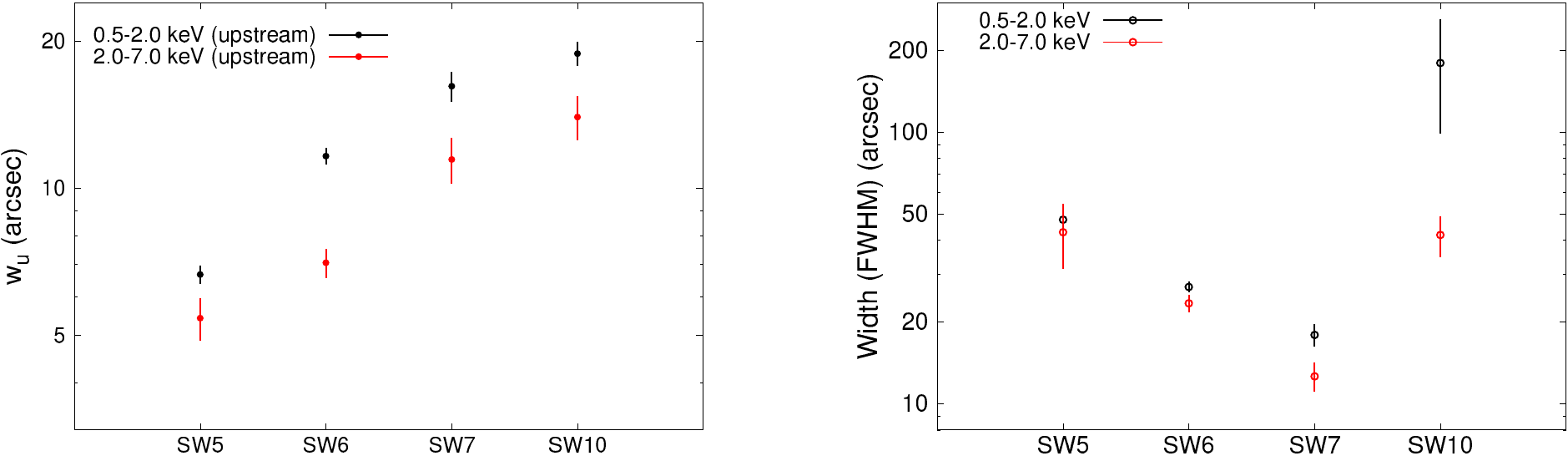}
\caption{{Upstream width parameter $w_{\rm u}$ and} filament widths in FWHM of the non-thermal filaments in the 0.5--2.0 keV and 2.0--7.0 keV energy ranges.
\label{fig-width}}
\end{figure*}

In order to investigate what determines the maximum energies of accelerated electrons seen in X-rays, the filament widths and their energy dependence are measured.
Here, we focus on the non-thermal-dominated filaments in 0.5--7.0~keV, SW5, SW6, SW7, and SW10.
We merge the images in 2001 and 2013 and use the merged one in this section.
We extract the flux radial profiles from the two energy ranges, 0.5--2.0~keV and 2.0--7.0 keV.
To model the profiles, we use a function
%\begin{linenomath}
\begin{equation}
\begin{cases}
F_{\rm u} (x) &= {C}\, \exp\left[-\frac{x-x_0}{w_{\rm u}}\right] \quad {(x \ge x_0)} \\
F_{\rm d} (x) &= {C}\, \exp\left[\frac{x-x_0}{w_{\rm d}}\right] \quad { (x < x_0)},
\end{cases}
\end{equation}
%\end{linenomath}
where the $F_{\rm u} (x)$ and $F_{\rm d} (x)$ are presumable upstream and downstream fluxes as functions of angular position $x$, respectively.
{The parameters $C$}, $x_0$, $w_{\rm u}$ and $w_{\rm d}$ are normalization parameter, angular position of the flux peak, and parameters to determine the widths in both regions, respectively.

The flux profiles in the soft and hard energy bands are presented in Figure~\ref{fig-profile-width}.
We fit the profile models to the data using the radial ranges where the filaments are bright compared to the background emission.
{We show the upstream width parameter $w_{\rm u}$ determined for the two energy ranges in the left panel of Figure~\ref{fig-width}.
Based on the derived parameters, we also calculate their filament widths in Full Width Half Maximum (FWHM), which are presented in the right panel of Figure~\ref{fig-width}.
Most cases show significantly narrower widths in the higher energies.}
%\footnote{{Including the broad flux profiles downstream the filaments SW5 and SW10 will not be appropriate to determine their widths. Thus we believe the upstream widths better represent the widths.}}
Thus, their maximum energies will be determined by synchrotron cooling, as in the case of Tycho's SNR \citep{tran15} and SN~1006 \citep{ressler14}, not by alternative processes such as the damping of downstream magnetic field (e.g., \citealt{pohl05}), which predicts filament widths independent of photon energy.

{We confirm that the energy dependence of the point spread function (PSF) at the positions of these filaments is $\lesssim 0\farcs5$ by comparing the PSFs at two representative energies, 1.0 and 4.0 keV (with the tool {\tt psfmap}; \citealt{allen04}). This small difference is because these filaments are located within $\sim 5'$ from the on-axis direction. Thus, the effect of the energy dependent PSF on the filament widths will be negligible.
Also, we check the energy dependence of the widths of the thermal-dominated filaments to examine possible systematics.
We find that the widths in the 0.5--0.7 and 0.7--1.2~keV energy ranges of the thermal-dominated filaments are consistent with each other.} \footnote{{For SW1, as an example, the filament widths in the lower and higher energy ranges are $42\pm6''$ and $49\pm3''$ in FWHM, respectively.}}

\subsection{Spectral Variations in the Downstream Regions of the Non-thermal Filaments}\label{sec-hardening}
Here we examine the spatial variation of the spectral shapes downstream the shocks for the non-thermal-dominated filaments.
{In this section, we focus on the filaments which are non-thermal-dominated in the 1.0--7.0~keV band, SW5, SW6, SW7, SW9, and SW10.} \footnote{{We find that the non-thermal parameters are constrained well if a spectrum is non-thermal-dominated above $\sim 1$~keV.}}
Spectral extraction regions are indicated in Figure~\ref{fig-image2}.
We merge the spectra in 2001 and 2013 to increase the statistics.
{Their proper motions are less than 30\% of the widths of the extraction regions.}
For the spectral modeling here, we use the model, Abs. (source emission) + (sky background) + PB.
The source emission is assumed to be {\tt powerlaw} + {\tt vpshock} for SW9 and {\tt powerlaw} for the others.
For the {\tt vpshock} model, only the normalization is treated as a free parameter whereas the other parameters are fixed to the best-fit values determined in Section~\ref{sec-fit}.

\begin{figure*}[htb!]
\centering
\includegraphics[width=16cm]{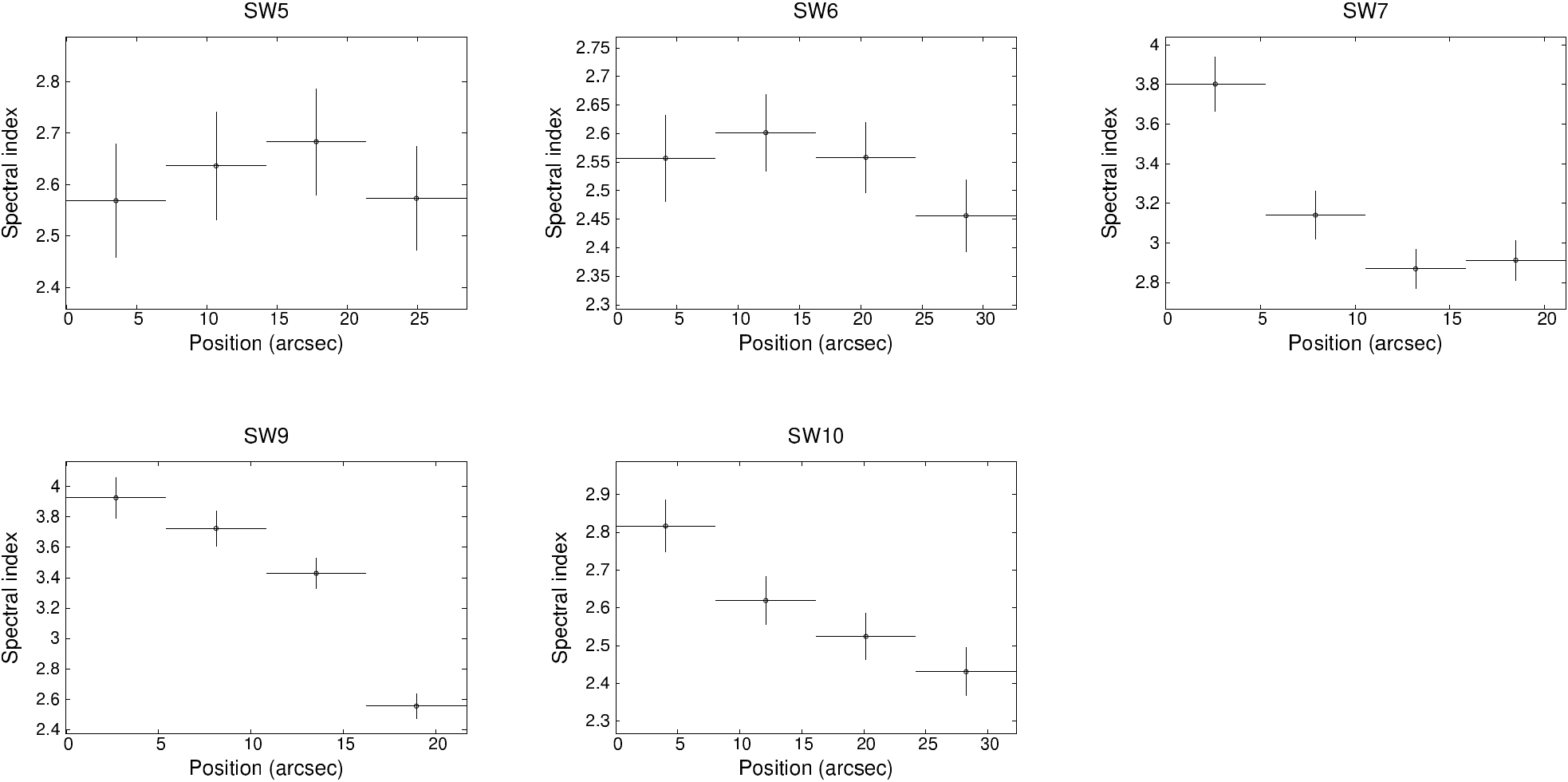}
\caption{Variation of power-law indices in the downstream regions toward the filaments for {SW5}, SW6, SW7, SW9, and SW10.
Increasing positions correspond to the directions of the arrows beside the regions in Figure~\ref{fig-image2}.
For each panel, the cross with the largest position is the data at the filament.
\label{fig-hardening}}
\end{figure*}

Resultant power-law indices are presented in Figure~\ref{fig-hardening}.
The SW7, SW9, and SW10 regions show spectral hardening in their shock downstream regions toward the shock fronts.
{This is another evidence that the synchrotron emission dominates the non-thermal component for these filaments (e.g., \citealt{katsuda10, kishishita13}).}
We confirm that the spectra extracted from all the regions are well explained with the {\tt powerlaw} model and thus thermal contamination will be negligible.

\subsection{Spectroscopy}
In order to extract thermal and non-thermal properties from individual filaments, we model here their energy spectra.

\subsubsection{Background Estimation}\label{sec-bgd}
Regarding the background (sky and particle-induced background), we estimate the sky background from a region outside the SNR shell (shown in Figure~\ref{fig-image}), and model the particle-induced background using the tool {\tt mkacispback} \citep{suzuki21b}.
We simultaneously model the spectra extracted from the background region for both 2001 and 2013 observations.
The spectral model for the background region is FE + Abs. (MWH + CXB) + PB, where the FE, Abs., MWH, CXB, and PB indicate the foreground emission \citep{yoshino09, kuntz00}, interstellar absorption, Milky Way halo (transabsorption emission) \citep{masui09, yoshino09, kuntz00}, cosmic X-ray background \citep{snowden92, kushino02, hickox06}, and particle-induced background model suited for our background region.

The FE component is described by the {\tt apec} model with the fixed temperature of 0.1~keV and metal abundances of solar values.
The emission measure, which is defined as (4$\pi${\it D}$^2$)$^{-1}$$\int${\it n}$_\mathrm{e}${\it n}$_\mathrm{H}${\it dV}~cm$^{-5}$, where {\it D}, {\it n}$_\mathrm{e}$. and {\it n}$_\mathrm{H}$ stand for distance, electron and hydrogen number densities, respectively, is treated as a free parameter.
The absorption column density of the Abs. component ({\tt tbabs} model) is fixed to $N_{\rm H} = 6.4 \times 10^{21}$ cm$^{-2}$ \citep{hi4pi16}.
The MWH component is also described with the {\tt apec} model with the fixed metal abundances of solar values and free temperature and emission measure.
The CXB component is described by the {\tt powerlaw} model with the fixed spectral index of 1.4 and the normalization corresponding to the flux of $6.38 \times 10^{-8}$ erg cm$^{-2}$ s$^{-1}$ str$^{-1}$ \citep{kushino02}.
As for the particle-induced background, the {\tt acispback} model predicts a lower flux than the observation in the $\sim 2$--7~keV band only in 2001 \citep{suzuki21b}.
Thus, we apply an additional {\tt powerlaw} model to {\tt acispback} only for the spectra in 2001.

\begin{figure}[htb!]
\centering
\includegraphics[width=8cm]{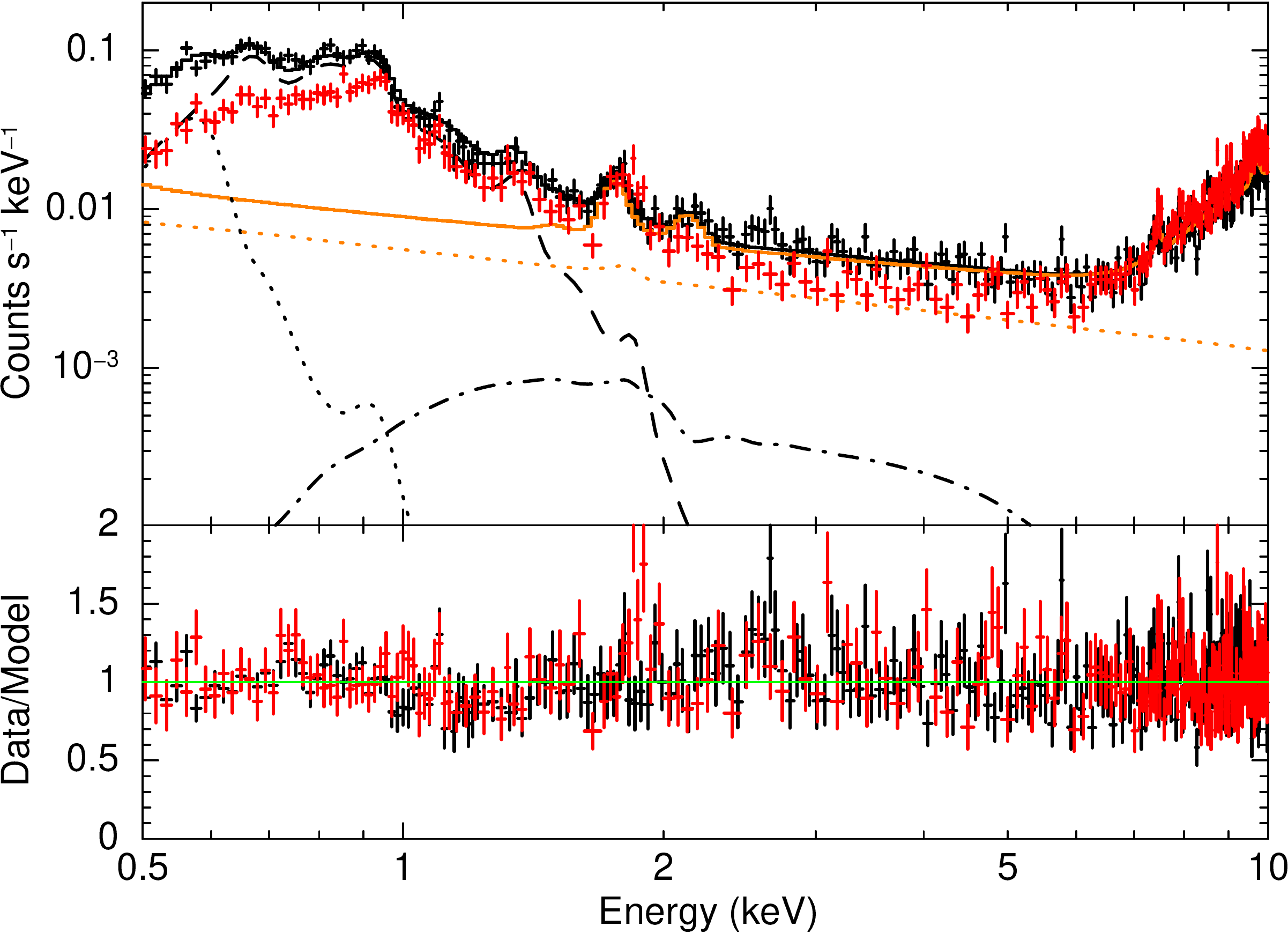}
\caption{Energy spectra and best-fit models for the background region.
The black and red crosses represent the data in 2001 and 2013, respectively.
The model spectra only for the data in 2001 are overplotted with lines.
The black solid line represents the entire model.
The black dotted, dashed, and dash-dotted lines show the FE, MWH, and CXB spectral models, respectively.
The orange dotted and solid lines represent the additional {\tt powerlaw} model to the PB component and the entire PB spectral model, respectively.
\label{fig-bgd}}
\end{figure}

The spectral fitting results are shown in Figure~\ref{fig-bgd}. The best-fit spectral parameters are summarized in Table~\ref{tab-bgd}.
The spectral parameters are consistent with \cite{kuntz00} and \cite{yoshino09}.
{Note that the visible flux decrease from 2001 to 2013 at the energies below 1~keV is due to the increased contamination on the sensor surface.}

\begin{table}[htb!]
%\fontsize{8}{9}\selectfont
\centering
\caption{Best-fit spectral parameters for the background region
\label{tab-bgd}}
\begin{tabular}{l l l }
\hline\hline
Model & Parameter & Value \\
\hline
Abs. & $N_{\rm H}$ ($10^{22}$ cm$^{-2}$) & 0.64 (fixed) \\
FE  &  $kT$ (keV)  & 0.1 (fixed) \\
 &  Abundance (solar)  & 1 (fixed) \\
 &  EM\tablenotemark{a} & $(9.7 \pm  1.2) \times 10^{-1}$  \\
MWH & $kT$ (keV) & $0.18 \pm 0.03$ \\
 &  Abundance (solar)  & 1 (fixed) \\
 & EM\tablenotemark{a} & $4.5 \pm 0.3$ \\
CXB & Photon index & 1.4 (fixed) \\
 & Normalization\tablenotemark{b} & $4.32 \times 10^{-6}$ (fixed) \\
\hline
\end{tabular}

\tablenotetext{a}{Emission measure, 10$^{-10}$(4$\pi${\it D}$^2$)$^{-1}$$\int${\it n}$_\mathrm{e}${\it n}$_\mathrm{H}${\it dV}~cm$^{-5}$, where {\it D}, {\it n}$_\mathrm{e}$. and {\it n}$_\mathrm{H}$ stand for distance (cm), electron and hydrogen number densities (cm$^{-3}$), respectively.}
\tablenotemark{b}{Normalization of the power-law model in units of cm$^{-2}$ s$^{-1}$ keV$^{-1}$ at 1 keV.}
\end{table}

\subsubsection{Spectral Modeling for Filaments}\label{sec-fit}
\begin{figure*}[htb!]
\centering
\includegraphics[width=16cm]{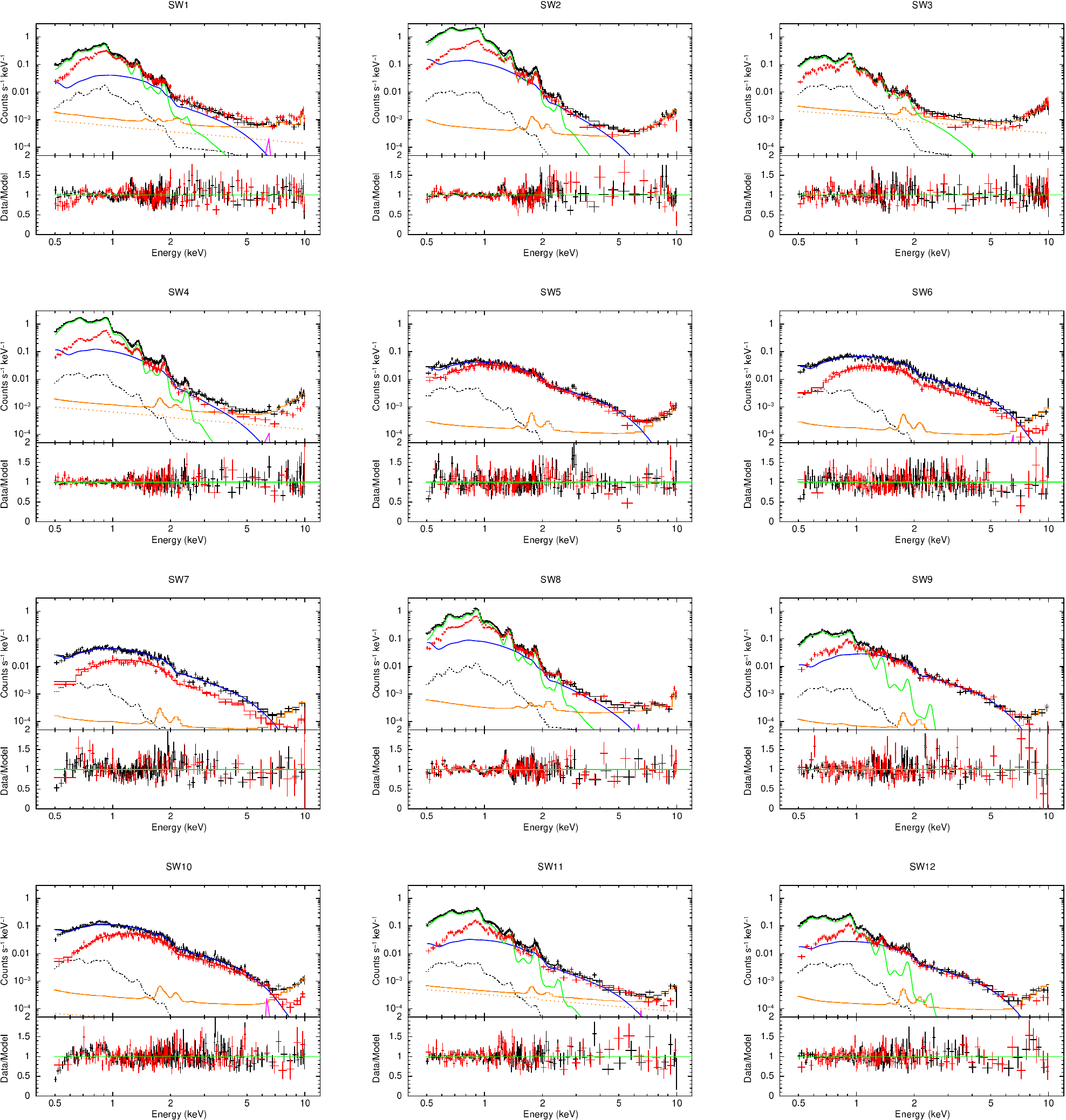}
\caption{Spectral modeling results for individual filaments. Black and red crosses are the observations in 2001 and 2013, respectively.
In each panel, the upper and lower panels show the flux and data-to-model ratio, respectively.
The best-fit model spectra only for the data in 2001 are overplotted with lines.
The black solid lines represent the entire models.
The green, blue, and magenta solid lines represent the thermal ({\tt vpshock}), non-thermal ({\tt powerlaw}), and Gaussian components, respectively.
The orange dotted and solid lines show the additional {\tt powerlaw} models to the PB component and the entire PB models, respectively.
The black dotted lines show the sky background components.
\label{fig-spectra}}
\end{figure*}

As the source emission, we apply the {\tt vpshock} (non-equilibrium ionizing plasma) model to the thermal, and the {\tt powerlaw} model to the non-thermal components following \cite{tsubone17}.
In addition, a Gaussian with the fixed energy centroid of 6.4~keV and width of zero is added for all the regions \citep{yamaguchi11, tsubone17}.
The spectral model is described as Abs. (source emission) + (sky background) + PB.
The source emission is {\tt powerlaw} for SW5, SW6, SW7, and SW10 because thermal components are found to be unnecessary.
The model {\tt vpshock} + {\tt powerlaw} is applied for the other regions.
{We note that the regions SW9 and SW12 are intermediate cases between the non-thermal- and thermal-dominated ones.} 
The absorption column density of the Abs. component ({\tt tbabs} model) is treated as a free parameter.
For {\tt vpshock}, the temperature, ionization timescale ($n_{\rm e} t$), emission measure, and the metal abundances of O, Ne, Mg, Si, and S are treated as free parameters.
The other metal abundances are fixed to solar values. All the parameters of {\tt vpshock} are tied between 2001 and 2013.
The {\tt powerlaw} normalization and index are treated as free parameters and are basically tied between 2001 and 2013. Only for SW5, SW6, SW7, and SW10, both parameters are untied between 2001 and 2013.
The sky background component is the same as that determined in Section~\ref{sec-bgd}. Their overall normalizations are scaled by the area ratio between the source regions and background region.
The PB component for the spectra in 2001 is composed of {\tt acipback} plus {\tt powerlaw} models, whereas that for 2013 is a simple {\tt acispback} model, which is the same treatment as that for the background estimation.

The spectra and best-fit models are presented in Figure~\ref{fig-spectra}.
The resultant spectral parameters are summarized in Table~\ref{tab-spectra}.
The power-law fluxes of SW5, SW6, SW7, and SW10 all show significant decrease from 2001 to 2013. Similarly, the power-law indices increase (spectra soften) from 2001 to 2013.
The absorption column densities, temperatures, metal abundances, and ionization timescales are roughly consistent with those presented in \cite{tsubone17}.
From the plasma densities inferred from the emission measures ($\sim 1$~cm$^{-3}$) and ionization timescales ($5\text{--}13 \times 10^{10}$~s~cm$^{-3}$), we confirm\footnote{With an assumption that our analysis regions are cuboids.} that the elapsed times after the shock heating are comparable to the remnant age ($\sim 2000$~yr).
The SW3 region may show smaller elapsed time of $\sim 600$~yr. This is consistent with the fact that this region is more distant from the remnant center than the other filaments and will be less affected by the projection effect, which mixes the newly heated and older regions.
{We also search for possible variations of the non-thermal parameters in the thermal-dominated cases where the spectral model fits the data relatively well (e.g., SW9 and SW12), and find consistent parameters in 2001 and 2013.} \footnote{{For example, the power-law fluxes of SW9 in the 0.5--7.0~keV band in 2001 and 2013 are $3.6\text{--}4.5 \times 10^{-13}$~erg and $4.4\text{--}5.5 \times 10^{-13}$~erg, respectively.}}

\begin{splitdeluxetable*}{l l l l B l l l l l l l l l}[htb!]
%\fontsize{8}{9}\selectfont
\tablewidth{0pt}
\tabletypesize{\scriptsize}
\tablecaption{Best-fit spectral-model parameters for the filament regions
\label{tab-spectra}}
\tablehead{
\colhead{Name} & \colhead{Abs.} & \multicolumn{2}{c}{{\tt powerlaw}} & \multicolumn{7}{c}{{\tt vpshock}} & \colhead{Gaussian} & \colhead{C-stat (d.o.f.)} \\
\cline{3-4} \cline{5-11}
\colhead{} & \colhead{$N_{\rm H}$ (10$^{22}$ cm$^{-2}$)} & \colhead{$\log_{10}$(Flux (erg~cm$^{-2}$~s$^{-1}$)) \tablenotemark{a}} & \colhead{Photon index} & \colhead{$kT$ (keV)} & \colhead{$n_{\rm e} t$ (10$^{10}$~cm$^{-3}$~s$^{-1}$)} 
& \colhead{EM \tablenotemark{b}} & \colhead{O (solar)} & \colhead{Ne (solar)} & \colhead{Mg (solar)} & \colhead{Si (solar)} & \colhead{${\rm Norm}$\tablenotemark{c}} & \colhead{} \\
}
\startdata
SW1 & $ 0.23 \pm 0.02 $ & $-12.21 \pm 0.02 $ & $3.14 \pm 0.12 $ & $ 0.60 \pm 0.04 $ & $ 4.9 \pm 0.9 $ & $ 6.9 \pm 1.0 $ & $ 0.74 \pm 0.05 $ & $ 1.43 \pm 0.09 $ & $ 1.31 \pm 0.09 $ & $ 1.26 \pm 0.14 $ & $ 9.3 \pm 4.5 $ & 1603.64 \, (1288) \\
SW2 & $ 0.34 \pm 0.02 $ & $-11.89 \pm 0.02 $ & $4.31 \pm 0.12 $ & $ 0.41 \pm 0.02 $ & $ 12.5 \pm 1.5 $ & $ 25.8 \pm 2.9 $ & $ 1.02 \pm 0.04 $ & $ 2.11 \pm 0.07 $ & $ 1.44 \pm 0.06 $ & $ 1.62 \pm 0.09 $ & $ 2.4 \pm 2.0 $ & 1798.70 \, (1288) \\
SW3 & $ 0.44 \pm 0.02 $ & \nodata & \nodata & $ 0.72 \pm 0.07 $ & $ 0.97 \pm 0.05 $ & $ 4.2 \pm 0.9 $ & $ 0.78 \pm 0.07 $ & $ 1.23 \pm 0.10 $ & $ 0.79 \pm 0.09 $ & $ 1.00 \pm 0.22 $ & $ < 3.5 $ & 1373.35 \, (1290) \\
SW4 & $ 0.43 \pm 0.02 $ & $-11.89 \pm 0.02 $& $4.18 \pm 0.12 $ & $ 0.38 \pm 0.02 $ & $ 7.7 \pm 1.0 $ & $ 25.2 \pm 4.4 $ & $ 1.59 \pm 0.08 $ & $ 2.49 \pm 0.12 $ & $ 1.45 \pm 0.08 $ & $ 2.28 \pm 0.16 $ & $ 3.3 \pm 2.3 $ & 1468.98 \, (1287) \\
SW5 & $ 0.26 \pm 0.01 $ & $-12.32 \pm 0.01 \, (-12.34 \pm 0.01$) & $2.59 \pm 0.07 \, (2.81 \pm 0.06$) & \nodata & \nodata & \nodata & \nodata & \nodata & \nodata & \nodata & $ 1.2 \pm 1.5 $ & 1405.35 \, (1293) \\
SW6 & $ 0.33 \pm 0.01 $ & $-12.04 \pm 0.01 \, (-12.10 \pm 0.01$) & $2.46 \pm 0.05 \, (2.66 \pm 0.05$) & \nodata & \nodata & \nodata & \nodata & \nodata & \nodata & \nodata & $ 2.4 \pm 2.1 $ & 1315.36 \, (1293) \\
SW7 & $ 0.32 \pm 0.02 $ & $-12.28 \pm 0.01 \, (-12.38 \pm 0.01$) & $2.82 \pm 0.08 \, (3.23 \pm 0.08$) & \nodata & \nodata & \nodata & \nodata & \nodata & \nodata & \nodata & $ < 1.8 $ & 1231.83 \, (1293) \\
SW8 & $ 0.27 \pm 0.01 $ & $-12.00 \pm 0.02 $ & $4.10 \pm 0.11 $ & $ 0.55 \pm 0.02 $ & $ 7.7 \pm 0.09 $ & $ 9.15 \pm 0.07 $ & $ 1.15 \pm 0.06 $ & $ 2.35 \pm 0.09 $ & $ 1.54 \pm 0.07 $ & $ 1.51 \pm 0.11 $ & $ 2.7 \pm 1.8 $ & 1799.19 \, (1287) \\
SW9 & $ 0.40 \pm 0.06 $ & $-12.34 \pm 0.01 $ & $2.66 \pm 0.08 $ & $ 0.36 \pm 0.02 $ & $ 9.5 \pm 3.6 $ & $ 2.6 \pm 1.1 $ & $ 1.52 \pm 0.18 $ & $ 2.67 \pm 0.41 $ & $ 1.85 \pm 0.36 $ & $ 1.90 \pm 0.86 $ & $ 1.4 \pm 1.2 $ & 1211.74 \, (1288) \\
SW10 & $ 0.25 \pm 0.01 $ & $-11.93 \pm 0.01 \, (-11.97 \pm 0.01$) & $2.57 \pm 0.05 \, (2.69 \pm 0.04$) & \nodata & \nodata & \nodata & \nodata & \nodata & \nodata & \nodata & $ 4.9 \pm 2.5 $ & 1468.25 \, (1293) \\
SW11 & $ 0.33 \pm 0.03 $ & $-12.52 \pm 0.02 $ & $2.98 \pm 0.12 $ & $ 0.50 \pm 0.04 $ & $ 7.3 \pm 1.7 $ & $ 3.0 \pm 0.6 $ & $ 1.00 \pm 0.08 $ & $ 2.16 \pm 0.16 $ & $ 1.36 \pm 0.13 $ & $ 1.95 \pm 0.25 $ & $ 3.9 \pm 1.7 $ & 1336.28 \, (1287) \\
SW12 & $ 0.28 \pm 0.05 $ & $-12.49 \pm 0.02 $ & $2.57 \pm 0.09 $ & $ 0.46 \pm 0.05 $ & $ 4.5 \pm 1.2 $ & $ 1.7 \pm 0.7 $ & $ 1.01 \pm 0.09 $ & $ 2.49 \pm 0.39 $ & $ 1.43 \pm 0.26 $ & $ 2.18 \pm 0.66 $ & $ < 2.0 $ & 1316.22 \, (1288) \\
\enddata

\tablenotetext{a}{Unabsorbed flux in the 0.5--7.0~keV energy range. For those with two values, the values in 2001 (2013) are shown.}
\tablenotetext{b}{Emission measure in the same units as that in Table~\ref{tab-bgd}}
\tablenotetext{c}{Total flux of the Gaussian model in units of 10$^{-7}$ photons~cm$^{-2}$~s$^{-1}$. The energy centroid of the Gaussian component is fixed to 6.4~keV.}

\end{splitdeluxetable*}

We also use the spectral model by \cite{zira07} (hereafter the ZA07 model) for the non-thermal component to discuss the acceleration efficiency.
The ZA07 model is described as
%\begin{linenomath}
\begin{equation}
\frac{dN}{dE} \propto\left(\frac{E}{E_{0}}\right)^{-2}\left[1+0.38\left(\frac{E}{E_{0}}\right)^{1 / 2}\right]^{11 / 4} \exp \left[-\left(\frac{E}{E_{0}}\right)^{1 / 2}\right],
\end{equation}
%\end{linenomath}
where $E$ and $E_0$ indicate photon energy and spectral turnover energy, respectively.
This model describes the synchrotron emission spectrum with several assumptions: loss-limited maximum energy, arbitrary energy dependence of the diffusion coefficient, shock compression ratio of four, and upstream-to-downstream magnetic field ratio of $\sqrt{11}$.
We replace the {\tt powerlaw} model in the spectral model set used above with the ZA07 model.
The free parameters for spectral modeling are the same as those in the case with the {\tt powerlaw} model except for the normalization and turnover $E_0$ of the ZA07 model.
The results are summarized in Table~\ref{tab-za07}.
One can see that the non-thermal-dominated filaments, SW5, SW6, SW7, and SW10, exhibit decreasing $E_0$ in time, which is the same tendency as our results with the {\tt powerlaw} model.

\begin{table}[htb!]
%\fontsize{8}{9}\selectfont
\centering
\caption{Best-fit parameters of the ZA07 model
\label{tab-za07}}
\begin{tabular}{l l }
\hline\hline
Name & $E_0$ (keV)\tablenotemark{a} \\
\hline
SW1 & $ 0.18 \pm 0.03 $ \\
SW2 & $ 0.041 \pm 0.004  $ \\
SW3 & \nodata \\
SW4 & $ 0.051 \pm 0.005 $ \\
SW5 & $ 0.42 \pm 0.06 \, (0.24 \pm 0.02) $ \\
SW6 & $ 2.35 \pm 0.76 \, (1.92 \pm 0.64) $ \\
SW7 & $ 0.26 \pm 0.03 \, (0.13 \pm 0.01) $ \\
SW8 & $ 0.08 \pm 0.01  $ \\
SW9 & $ 1.72 \pm 0.72  $ \\
SW10 & $ 0.45 \pm 0.04 \, (0.33 \pm 0.02) $ \\
SW11 & $ 0.24 \pm 0.05  $ \\
SW12 & $ 2.52 \pm 1.69  $ \\
\hline
\end{tabular}

\tablenotetext{a}{Values are tied between the data in 2001 and 2013 except for those for SW5, SW6, SW7, and SW10, for which the values in 2001 (2013) are shown.}
\end{table}

\section{Discussion} \label{sec-discussion}

\subsection{Magnetic field amplitude}\label{sec-mag}
\begin{figure}[htb!]
\centering
\includegraphics[width=8cm]{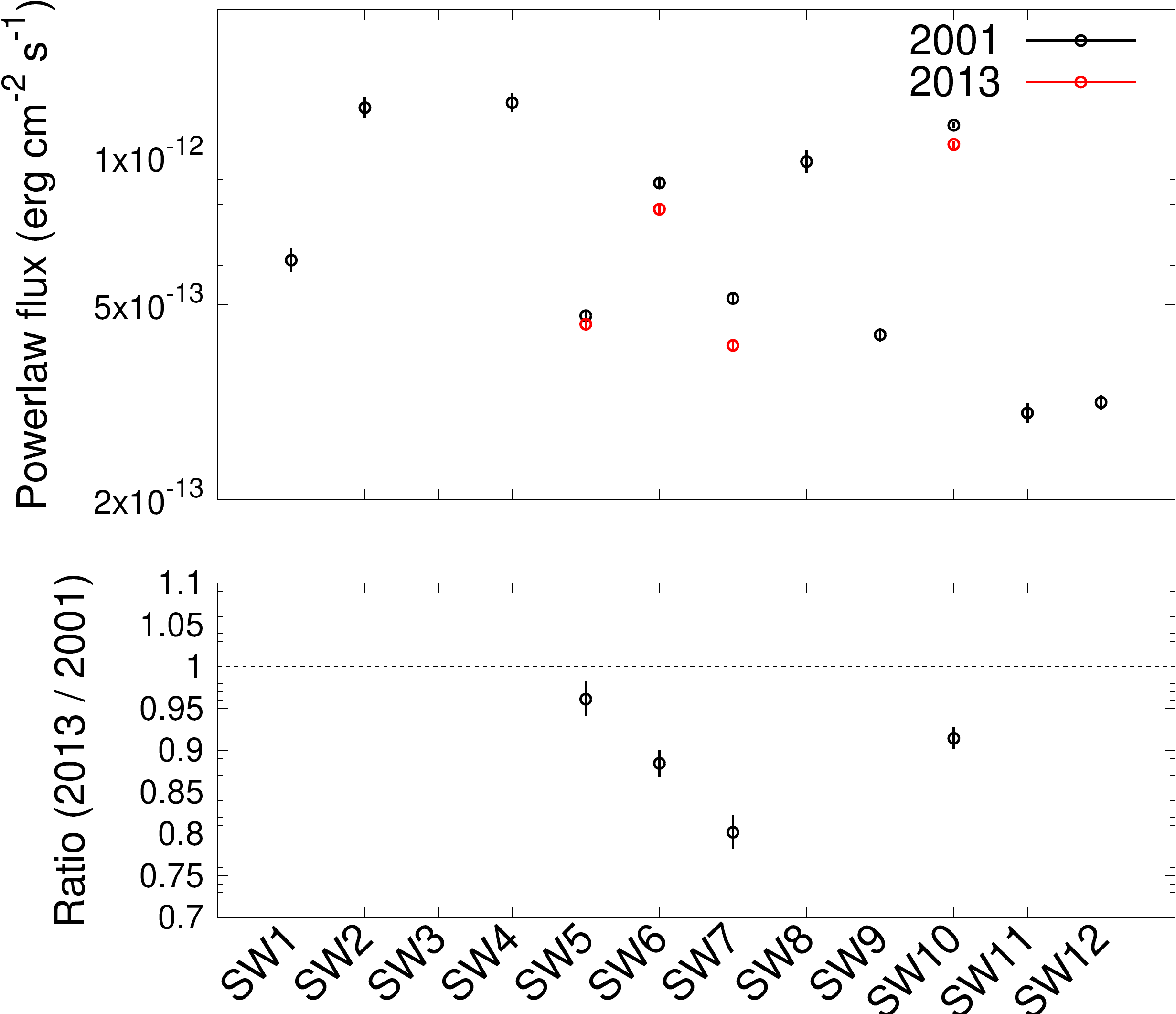}
\caption{Power-law fluxes of the individual filaments (top) and their 2013-to-2001 ratios (bottom; only for the non-thermal dominated filaments).
The flux is calculated for the 0.5--7.0~keV energy range.
\label{fig-PLflux}}
\end{figure}

We have found that the non-thermal-dominated filaments SW5, SW6, SW7, and SW10 showed flux decrease from 2001 to 2013. This is seen in Figure~\ref{fig-PLflux}.
The decrease rates are from $\sim 0.4\% \text{ yr}^{-1}$ (SW5) to $\sim 2\% \text{ yr}^{-1}$ (SW7).
Given that the non-thermal emissions are dominated by the synchrotron radiation, it is reasonable to attribute such flux decreases to the synchrotron cooling.
The cooling timescale is described as
%\begin{linenomath}
\begin{equation}
t_{\rm syn} = 50 \, {\rm yr} \, \left(\frac{E}{1~{\rm keV}}\right)^{-0.5} \left(\frac{B}{100~\mu{\rm G}}\right)^{-1.5},
\end{equation}
%\end{linenomath}
where $E$ and $B$ represent the synchrotron photon energy and magnetic field strength, respectively.
Thus, the flux decreases roughly require the field strengths of $\sim 30$--100~$\mu$G.
Such field strengths are similar to the previous estimates (24~$\mu$G by \citealt{vink06} based on filament widths; 35~$\mu$G by \citealt{helder12} based on filament widths; 14--20~$\mu$G by \citealt{yuan14} based on the broadband spectrum).

The magnetic field strength at a filament can also be estimated based on the filament width seen in X-rays.
As presented in \cite{vink06}, the magnetic field strength is simply determined by the filament width $w$ at the target distance $d$ as
%\begin{linenomath}
\begin{equation}\label{eq-width}
\begin{aligned}
B &= 54~\mu{\rm G} \, (w/0.1~{\rm pc})^{-2/3} \\
&\approx 54~\mu{\rm G} \, (w/7.3'')^{-2/3} (d/2.8~{\rm kpc})^{-2/3},
\end{aligned}
\end{equation}
%\end{linenomath}
if the energy range for the width estimation is near the roll-off energy of the synchrotron emission.
We thus obtain magnetic field strengths of $\sim 20$--50~$\mu$G for SW5, SW6, SW7, and SW10 based on Figure~\ref{fig-width}.
These estimates are also consistent with the previous estimates \citep{vink06, helder12, yuan14}.

\subsection{Magnetic-field turbulence level}
\begin{figure}[htb!]
\centering
\includegraphics[width=8cm]{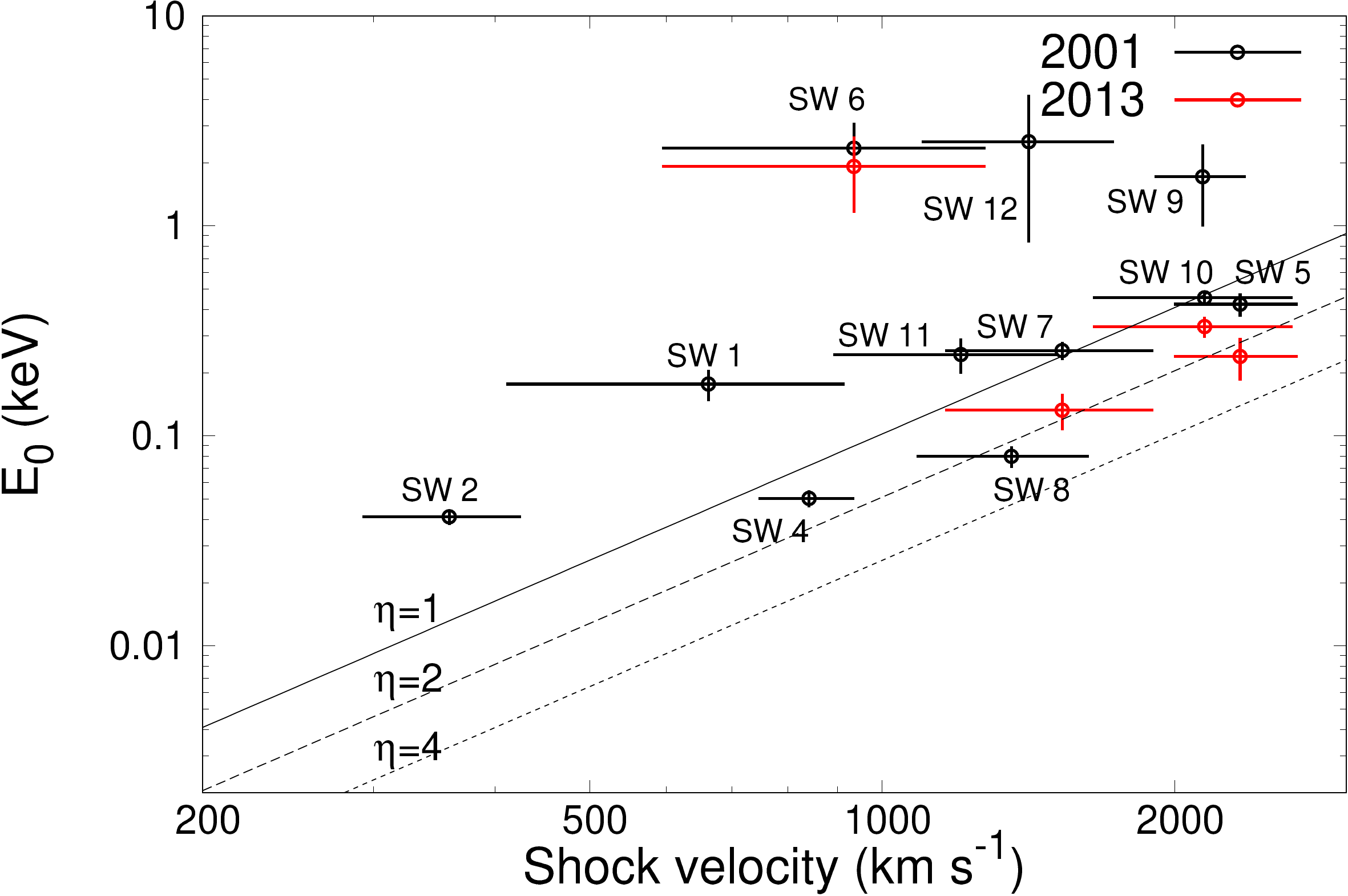}
\caption{Plots of the cutoff energy of the ZA07 model ($E_0$) against the shock velocity (proper motion velocity is substituted).
A distance of 2.8~kpc is assumed.
The black solid, dashed, and dotted lines indicate the ZA07 model with different gyrofactors ($\eta$).
\label{fig-e0}}
\end{figure}

Based on the shock velocity ($v_{\rm sh}$) and turnover energy of the ZA07 model ($E_0$), one can estimate the gyrofactor ($\eta = (B/\delta B)^2$, where $B$ and $\delta B$ are the magnetic field amplitude and fluctuation, respectively).
\cite{zira07} derived $E_0$ as a function of $\eta$ and $v_{\rm sh}$ as
%\begin{linenomath}
\begin{equation}\label{eq-eta}
E_0 = 0.24~{\rm keV} \, \eta^{-1} (v_{\rm sh}/2000~{\rm km~s^{-1}})^2.
\end{equation}
%\end{linenomath}
Figure~\ref{fig-e0} exhibits the plots of $E_0$ against the shock velocity, for which the proper motion velocity is simply substituted.
Compared to the ``$\eta = \text{const.}$'' lines, SW3, SW4, SW5, SW7, SW8, SW10, SW11, and SW12 are consistent with $\eta = 1$--4, whereas SW1, SW2, SW6, and SW9 have $\eta < 1$.
Such small values of $\eta$ indicate the existence of highly amplified magnetic turbulence.
This is similar to the situations of SN~1006 \citep{ressler14} and RX~J1713.7$-$3946 \citep{tsuji19}.
{\cite{dickel01} found a low degree of polarization in the RCW~86 SW region and proposed a Faraday depolarization scenario. The high level of magnetic turbulence suggested from our results can be an alternative for this low degree of polarization.}

Given that RCW~86 is thought to have begun to interact with the cavity wall recently \citep{williams11}, such small $\eta$ values might indicate that the shocks have been decelerated just recently, whereas the accelerated particles remain as they were before the interaction. This would be reasonable if the synchrotron cooling timescales we evaluated ($\gtrsim 50$--200~yr) are longer than the deceleration timescales of the shocks.
{We basically believe that the projection effect on the proper motions will be small ($\lesssim 15\%$; see Section~\ref{sec-prop}), but it still possibly contributes to such small $\eta$ values because of the complicated shock structure.}
It is also possible that these small $\eta$ values are due to the uncertainties in the distance to RCW~86.
A factor of two larger distance would result in $\eta \gtrsim 1$ for most of the filaments.
Note that larger distances would, on the other hand, make the magnetic-field estimates based on the filament widths be in tension with those from the flux decrease (Section~\ref{sec-mag}).
%The SW6 filament will be associated with the highest maximum energy among our analysis regions, as indicated from Figure~\ref{}.

\subsection{Nature of the inward-moving filaments}\label{sec-inward}
We have found that the SW7 and SW10 filaments are moving inward.
The spectral softening toward the downstream regions is also confirmed (Section~\ref{sec-hardening}), which supports their inward movements.
Here we discuss their nature: reverse or reflected shocks.

Assuming that they are reverse shocks, we can estimate their shock velocities in the unshocked-ejecta frame with an assumption of the explosion center coordinates.
The explosion center is substituted with the geometrical center, and is assumed to be $(l, b) = (315\fdg4068, -2\fdg3151)$ by fitting the radio shell with a circle by eye.
The ejecta velocities in the laboratory frame ($v_{\rm ej}$) and the estimated shock velocities ($v_{\rm rev}$) are summarized in Table~\ref{tab-rshock}.
These velocities would be too large considering their values of $E_0$, i.e., they would require much larger values of the gyrofactor $\eta$ compared to the other filaments (Figure~\ref{fig-e0}).
Given that SW7 and SW10 are associated with rather hard X-ray emission, it would be unreasonable if they were outliers in Figure~\ref{fig-e0} with particularly large $\eta$ values.\footnote{Note that the large $\eta$ values of $\sim 20$ suggested in the reverse-shock scenario for SW7 and SW10 are not unreasonable themselves, because similar values of $\eta$ have been found in other SNRs with similar ages \citep{tsuji21}.}
Besides, the presumable reverse-shock velocities of $\sim 10000$~km s$^{-1}$ are even larger than the shock velocities inferred for the inward-moving filaments observed in Cassiopeia~A (e.g., \citealt{sato18}).

On the other hand, if we assume that the inward-moving filaments are reflected shocks, we cannot simply estimate their shock velocities without certain assumptions (e.g., \citealt{truelove99}).
In this case, the reflected shocks should be moving in the shocked ejecta.
If the ejecta would have been decelerated significantly (e.g., by the cavity wall as suggested by \citealt{williams11} or dense clouds as suggested by \citealt{sano19b}), the shock velocities with respect to the shocked ejecta would be as small as those of the other filaments (Figure~\ref{fig-e0}).
Thus, we propose the reflected-shock scenario as the most likely nature of the two inward-moving filaments.

\begin{table}[htb!]
%\fontsize{7}{9}\selectfont
\centering
\caption{Shock velocities in the reverse shock scenario} \label{tab-rshock}
\begin{threeparttable}
\begin{tabular}{l l l r}\hline\hline
Name & Radius ($'$)\tablenotemark{a} & $v_{\rm ej}$ (km~s$^{-1}$)\tablenotemark{b} & $v_{\rm rev}$ (km~s$^{-1}$)\tablenotemark{c} \\\hline
SW5 & 22.8826 & 9668 & $7300 \pm 700$ \\
SW6 & 17.6610 & 7462 & $6500 \pm 700$ \\
SW7 & 18.4815 & 7809 & $9300 \pm 900$ \\
SW10 & 18.1032 & 7649 & $9800 \pm 1000$ \\
\hline
\end{tabular}

\tablenotetext{a} {Filament position with respect to the geometrical center $(l, b) = (315\fdg4068, -2\fdg3151)$}
\tablenotetext{b} {Fast-moving ejecta velocity in the laboratory frame}
\tablenotetext{c} {Shock velocity in the ejecta frame}

\end{threeparttable}
\end{table}

\subsection{Maximum energies of accelerated protons}
As we have obtained several parameters related to particle acceleration, the age $t$, shock velocity $v_{\rm sh}$, magnetic field strength $B$, and gyrofactor $\eta$, we derive here the maximum energies of particles.
Those of protons are of particular interest. With an assumption that the proton maximum energy ($E_{\rm max, p}$) is limited by the acceleration time and that the diffusion coefficient is time-invariant, $E_{\rm max, p}$ can be written as (e.g., \citealt{reynolds08, yamazaki14})
%\begin{linenomath}
\begin{equation}
E_{\rm max, p} \approx 200~{\rm TeV} \, \eta^{-1} \left(\frac{v_{\rm sh}}{2000 \, \text{km s$^{-1}$}}\right)^2 \left(\frac{B}{50 \, \text{$\mu$G}} \right) \left(\frac{t}{2 \, \text{kyr}} \right).
\end{equation}
%\end{linenomath}
If we substitute the parameters for SW7, $\eta \approx 1$, $v_{\rm sh} \approx 1600$~km~s$^{-1}$, and $B \sim 100~\mu$G (based on the non-thermal flux decrease), we obtain $E_{\rm max, p} \sim 210$~TeV.
Similarly, for SW5, SW6, and SW10, we obtain $E_{\rm max, p} \sim 140$, 130, and 240~TeV, respectively.
These estimates are consistent with the maximum energies evaluated from the gamma-ray spectrum assuming hadronic gamma-rays, $\approx 10\text{--}200$~TeV \citep{yuan14, zeng19, suzuki20b, suzuki22a}\footnote{We note that our estimates are also consistent with the leptonic scenario, where the proton maximum energy of $\gtrsim 20$~TeV is indicated.}.
Thus, our parameter estimates such as the magnetic field strength and turbulence level would be reasonable.
%Thus, our estimates may suggest that the gamma-rays are of hadronic origin, although they will not rule out the leptonic scenario, where the proton maximum energy of $\gtrsim 20$~TeV is indicated.

\subsection{Acceleration parameters and their dependence on environment}

\begin{figure*}[htb!]
\centering
\includegraphics[width=16cm]{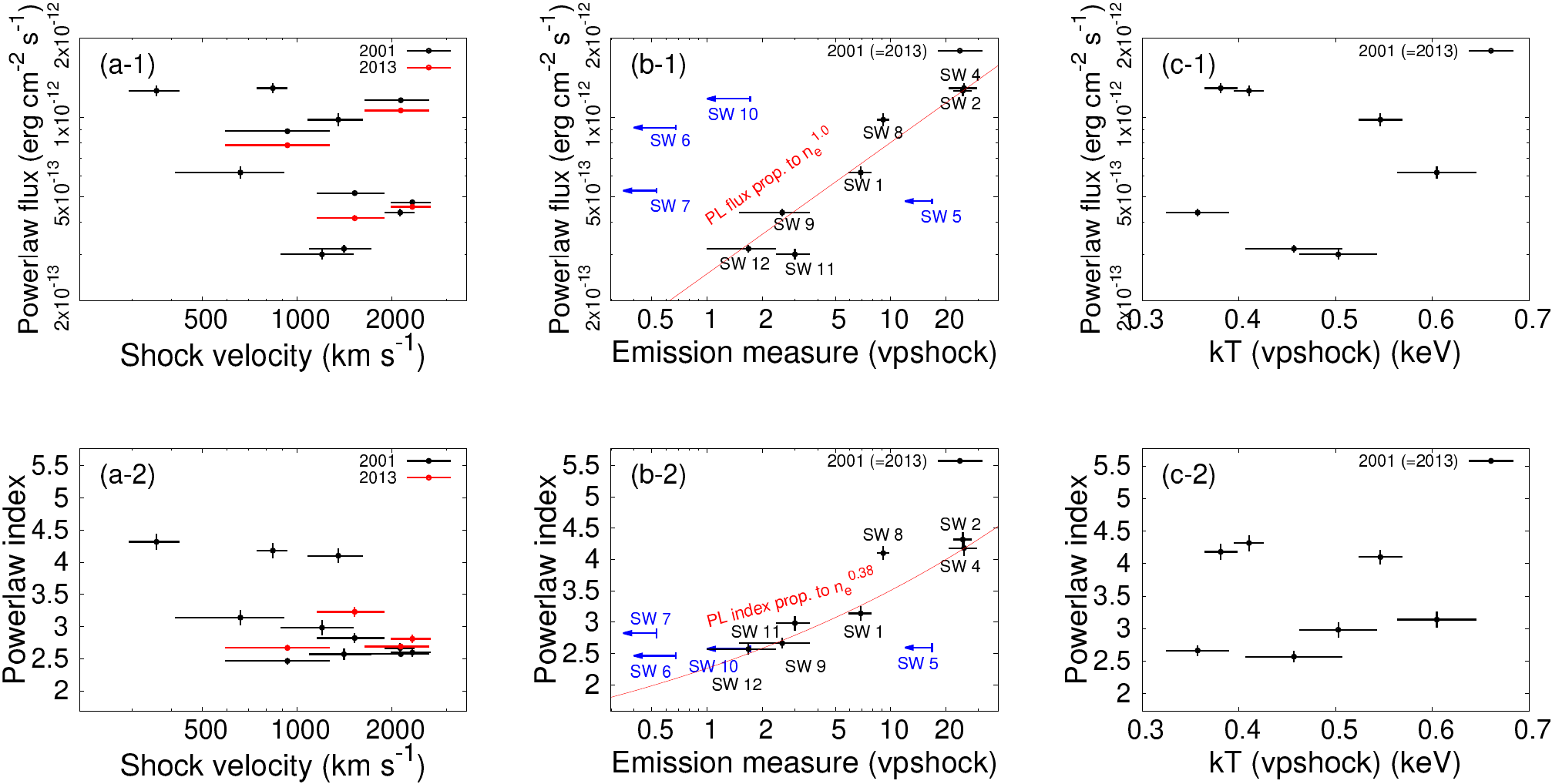}
\caption{Plots of the power-law flux (PL flux) and index (PL index) as functions of the shock velocity (proper motion velocity is substituted), emission measure ($\propto \text{density } (n_{\rm e})^2$), and electron temperature ($kT$).
The PL flux is calculated for 0.5--7.0~keV. {The EM is shown in the same units as Table~\ref{tab-bgd}.}
{The 95\% upper limits of the EMs of the non-thermal-dominated filaments are shown with the blue arrows in panels (b).}
The best-fit power-law functions for the plots of the PL flux and index over the emission measure are also shown with red lines.
\label{fig-PL}}
\end{figure*}

\begin{figure}[htb!]
\centering
\includegraphics[width=8cm]{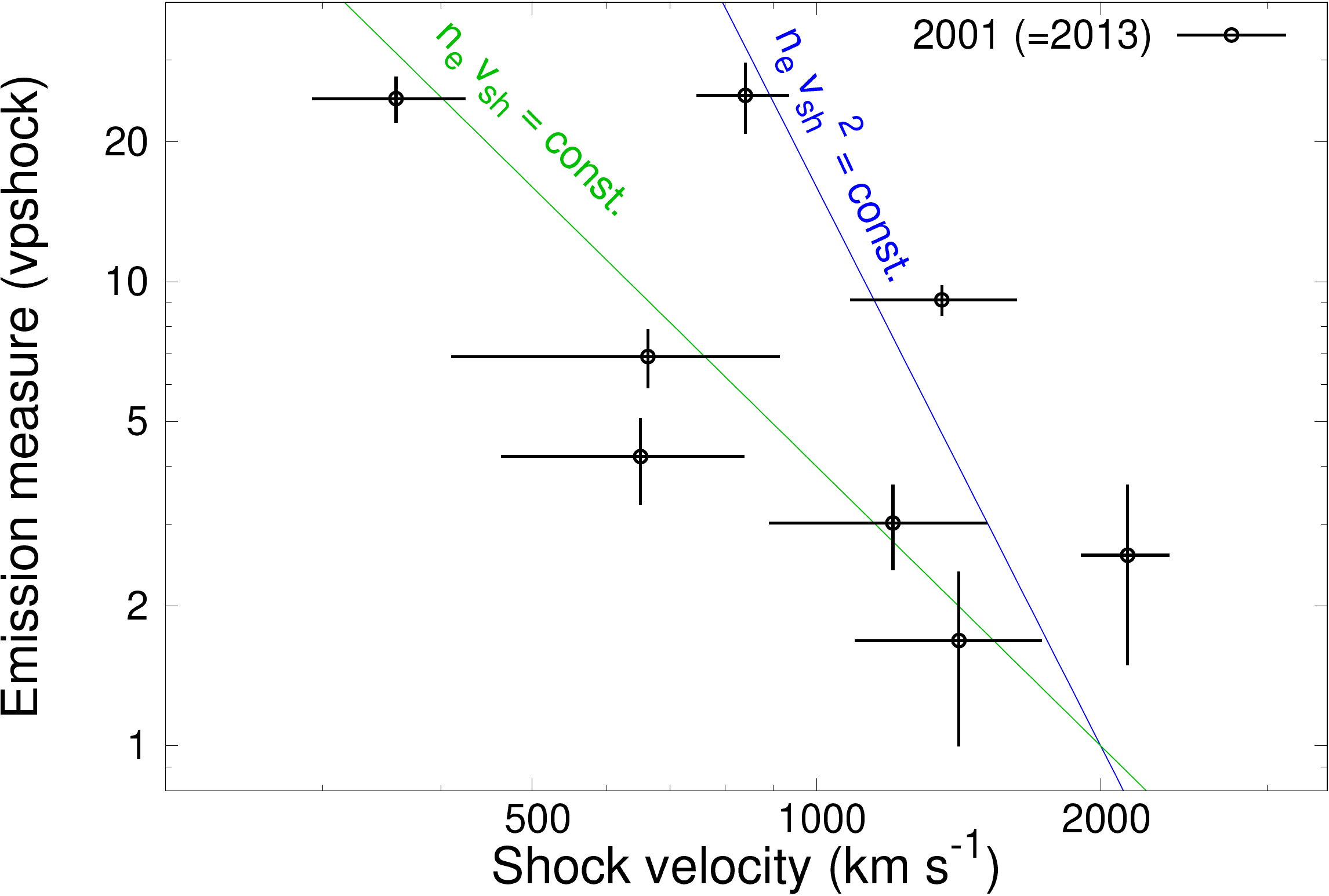}
\caption{Plots of the emission measure ($\propto n_{\rm e}^2$) as a function of the shock velocity ($v_{\rm sh}$; proper motion velocity is substituted).
A distance of 2.8~kpc is assumed. The EM is shown in the same units as Table~\ref{tab-bgd}.
The solid lines indicate two conditions, $n_{\rm e} v_{\rm sh} =  \text{const.}$ (green) and $n_{\rm e} v_{\rm sh}^2 =  \text{const.}$ (blue).
\label{fig-ne-vs-vsh}}
\end{figure}

\subsubsection{Parameter correlations}
Based on the proper motions and spectral parameters we have obtained, we discuss here their parameter correlations to see how the environmental parameters affect the acceleration processes.
Figure~\ref{fig-PL} shows the power-law parameters versus shock velocity, thermal emission measure, and electron temperature.
{Focusing on the filaments with both thermal and non-thermal parameters obtained (black crosses)}, the power-law parameters clearly depend on the emission measure ($\propto n_{\rm e}^2 V$, where $n_{\rm e}$ and $V$ are plasma density and volume, respectively) and do not or only weakly on the shock velocity and post-shock electron temperature.
If we evaluate the correlation between the downstream plasma density and power-law parameters with a power-law function, we obtain $\text{(power-law flux)} \propto n_{\rm e}^{1.0 \pm 0.2}$ and $\text{(power-law index)} \propto n_{\rm e}^{0.38 \pm 0.10}$.
{Note that the non-thermal-dominated filaments, SW5, SW6, SW7, and SW10 are not included in these evaluations.}
Such correlations were indeed implied by \cite{tsubone17} using Suzaku by investigating the whole remnant, but are more clarified in this work with Chandra.
An increase in the non-thermal flux associated with high-density regions was also observed in Cassiopeia~A (\citealt{sato18, fraschetti18})
%It is interesting that the non-thermal parameters are almost simply determined by the plasma density, not or only weakly by the shock velocity.

As seen in Figure~\ref{fig-ne-vs-vsh}, the emission measure is negatively correlated with the shock velocity $v_{\rm sh}$.
This correlation appears to follow the function $n_{\rm e} v_{\rm sh} = \text{const.}$
A simple assumption of the constant ram pressure $n_{\rm e} v_{\rm sh}^2 = \text{const.}$ seems to be inapplicable.
This may reflect the inhomogeneity in the ambient density suggested by the complicated structure of the remnant and discussed in previous works \citep{williams11, tsubone17}.
We note that additional uncertainties of the shock velocity are expected in the moving direction and projection effect, which would be less than a factor of $\sim$ two.

\begin{figure}[htb!]
\centering
\includegraphics[width=8cm]{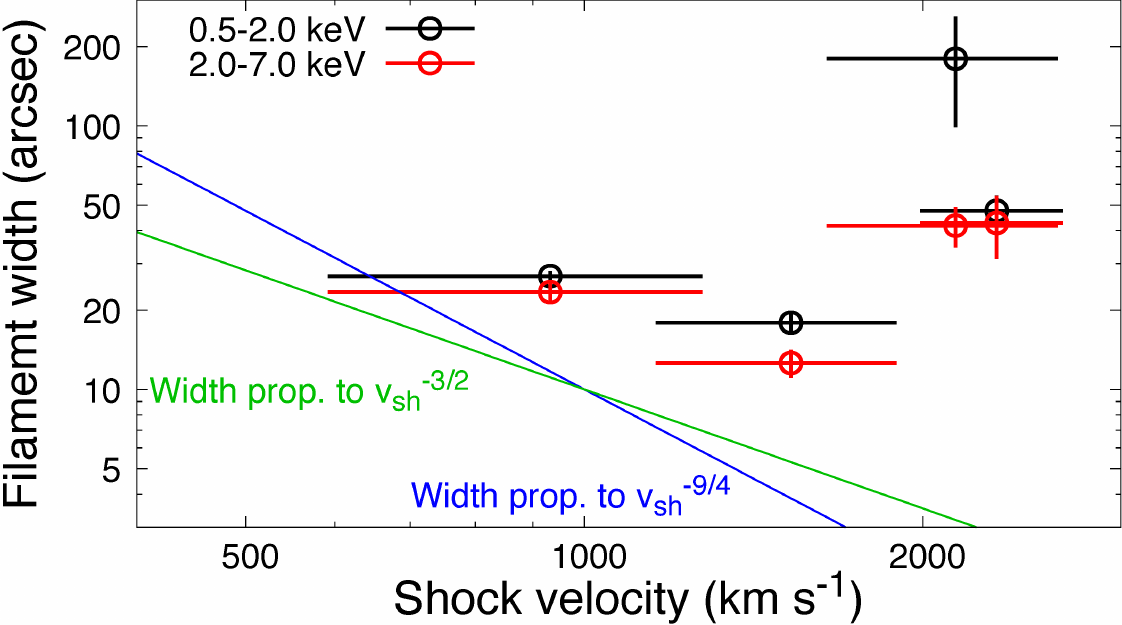}
\caption{Plots of the filament width ($w$) against the shock velocity ($v_{\rm sh}$; proper motion velocity is substituted).
A distance of 2.8~kpc is assumed.
The solid lines indicate two functions, $w \propto v_{\rm sh}^{-3/2}$ (green) and $w \propto v_{\rm sh}^{-9/4}$ (blue).
\label{fig-b-width}}
\end{figure}

Certain correlations among magnetic field strength $B$, density $n_{\rm e}$, and shock velocity $v_{\rm sh}$ are expected due to the Bell instability \citep{bell04}, and were in fact confirmed from observations of several young SNRs \citep{volk05, vink06b, helder12, vink17}: $B^2 \propto n_{\rm e} v_{\rm sh}^3$ or $B^2 \propto n_{\rm e} v_{\rm sh}^2$.
If we plot the filament width ($\propto B^{-3/2}$ by Eq.~\ref{eq-width}) over the shock velocity for SW5, SW6, SW7, and SW10, we obtain Figure~\ref{fig-b-width}.
As can be seen, the data do not follow the relations expected above.

{In order to compare the parameter correlations of the non-thermal-dominated filaments to those of the thermal-dominated ones, we evaluate the upper limits of the EMs for the non-thermal-dominated filaments (SW5, SW6, SW7, SW9, and SW10).
We repeat the spectral modeling described in Section~\ref{sec-fit} with an additional ionizing plasma model with various electron temperatures (0.3--0.8~keV) and ionization timescales ($10^{10\text{--}11}$ s~cm$^{-3}$) with the metal abundances fixed to solar, and derive the upper limit of the EM. \footnote{{The two parameter ranges are selected based on the values determined for the thermal-dominated filaments (Table~\ref{tab-spectra}).}}
The resulting upper limits of the EMs versus the power-law parameters are shown in Figure~\ref{fig-PL} (b).
The non-thermal-dominated filaments located at the inner regions, SW6, SW7, and SW10, do not follow the same tendency as that of the other filaments (shown with the red solid lines).
}

We have also investigated the dependence of the non-thermal parameters on the shock obliquity (angle between the shock normal and background magnetic field).
The background field direction is estimated from the starlight polarization at similar distances to RCW~86 (see Appendix~\ref{sec-other} for more details).
The field direction around RCW~86 is found to be nearly parallel to the Galactic plane.
We find no correlation between the non-thermal parameters and the shock obliquity as can be seen in Figure~\ref{fig-pl-obliquity}.
%This probably reflects the uncertainties associated with the moving directions of the filaments, as suggested also by the plots of $n_{\rm e}$ versus $v_{\rm sh}$.

\subsubsection{Scenario to explain the measurements: shock-cloud interaction}
%Considering the parameter correlations above, we propose here a shock-cloud interaction scenario to explain them.
We propose here a shock-cloud interaction scenario to explain the parameter correlations we have obtained above.
As the RCW~86 SW region is interacting with dense atomic and molecular clouds \citep{sano17, sano19b}, X-ray emission is expected to trace such an interaction.
The shock-cloud interaction will slow down the shocks and will damp the magnetic turbulence inside dense clumps but amplify the magnetic turbulence around them \citep{inoue12, fraschetti13}.
If our analysis has resolved such clumpy structures, we expect harder synchrotron X-rays for lower-density regions due to higher maximum acceleration energies, which matches the measured trend (Figure~\ref{fig-PL} (b)).
We note that our spectral extraction regions have a size of $\sim 0.2$--0.4~pc at a distance of 2.8~kpc, and the shock crossing time is estimated as $\sim 100$--1000~yr. Since the cooling timescale derived for the non-thermal-dominated filaments of $\sim 50$--200~yr is similar to the shock crossing timescale, it will be reasonable to assume that the spatial extent of clumps are similar to our region sizes.
The fact that the power-law flux is proportional to the density (Figure~\ref{fig-PL} (b-1)) can be understood as more enhanced non-thermal particle density in higher-density regions.
%The shock structures are disturbed by the interaction, leading to large dispersions in the power-law parameters at the similar shock velocities (Figure~\ref{fig-PL}), as well as at the similar shock obliquities (Figure~\ref{fig-pl-obliquity}).
The shock-cloud interaction leads highly spatially inhomogeneous magnetic fields, and so the power-law parameters show large scatters even at similar shock velocities (Figure~\ref{fig-PL} (a)).
However, interestingly, the plasma density seems to remain as a good tracer of the modification of the acceleration conditions even after the interaction (Figure~\ref{fig-PL} (b)).

%The reason of the apparent inconsistency of our estimates of the magnetic field strengths and shock velocities with the Bell instability (Figure~\ref{fig-b-width}) is understandable if the shock-cloud interaction greatly modifies the local magnetic turbulence levels, even though the global field strengths for the whole remnant are controlled by the Bell instability according to a comparison to other SNRs \citep{vink06b, helder12}.
The reason for the apparent inconsistency of our estimates of the magnetic field strengths and shock velocities with the Bell instability (Figure~\ref{fig-b-width}) is understandable if the magnetic turbulence induced by the shock-cloud interaction determines local acceleration activity.
Note that, according to a comparison with other SNRs, the base magnetic field amplification level over the whole remnant seems to be controlled by the Bell instability \citep{vink06b, helder12}.
Thus, it is suggested that the local magnetic turbulence levels $\delta B/B$ are of great importance to understand the local acceleration conditions.

{In Figure~\ref{fig-PL} (b), the tendency of the inner non-thermal-dominated filaments (SW6, SW7, and SW10) differs from that of the outer filaments.
%This discrepancy will be due to different acceleration environments between the outer thermal-dominated and inner non-thermal-dominated regions.
If we assume that the outer filaments are newly interacting with dense gas and the inner ones are still in the wind-blown bubble, our results may suggest that the cavity region has different acceleration conditions from those of the interacting regions, which is naturally expected.}

As a conclusion, our findings suggest that the acceleration physics {at the outer filaments} of the RCW~86 SW region is governed by the ambient density, not or only weakly by the shock velocity and shock obliquity.
We find that the shock-cloud interaction scenario can explain the measurements consistently, although not yet in a quantitative manner.
Radio observations with high angular resolutions (e.g., Atacama Large Millimeter/submillimeter Array: ALMA) will greatly help test our scenario.

\section{Conclusion}
In this work, we studied the X-ray proper motions and spectral properties of the RCW~86 SW region.
The proper motion velocities were found to be $\sim 300$--2000~km~s$^{-1}$ at a distance of 2.8~kpc.
We found two inward-moving filaments. They were found to be non-thermal dominated and the spectral softening toward downstream were seen, which confirmed their inward movements.
It is likely that they are reflected shocks rather than reverse shocks.
Based on the X-ray spectroscopy, we evaluated thermal parameters such as the ambient density and temperature, and non-thermal parameters such as the power-law flux and index.
Also, based on the flux decrease of several non-thermal filaments, we were able to estimate the magnetic field amplitudes of $\sim 30$--100~$\mu$G.

Gathering the proper motion and X-ray properties, we then studied the parameter correlations.
We found that, {at the outer thermal-dominated filaments}, the non-thermal parameters were correlated with the ambient density as $\text{(power-law flux)} \propto n_{\rm e}^{1.0 \pm 0.2}$ and $\text{(power-law index)} \propto n_{\rm e}^{0.38 \pm 0.10}$, not or only weakly with the shock velocity and shock obliquity.
These indicate harder and fainter synchrotron emission for lower-density regions.
As an interpretation of the measured physical parameters, we propose the shock-cloud interaction scenario, where the locally enhanced magnetic turbulence levels ($\delta B/B$) have a great influence on the local acceleration conditions.
{The inner non-thermal-dominated filaments showed a different tendency from that of the outer filaments, which is understandable if the inner ones are still in the wind-blown bubble and have different acceleration conditions.}

\acknowledgments
We appreciate helpful suggestions provided by the anonymous referee, which has improved the paper significantly.
We are grateful to K. Kawabata for providing the software to extract the starlight polarization data.
This work was partially supported by JSPS grant Nos.~19J11069 and 21J00031 (HS), 20H00174, 21H01121 (SK), 19H01936,  21H04493 (TT), and 20H01944 (TI).
FF was supported, in part, by NASA through Chandra Theory Award Number $TM0-21001X$, issued by the Chandra X-ray Observatory Center, which is operated by the Smithsonian Astrophysical Observatory for and on behalf of NASA under contract NAS8-03060.

\vspace{5mm}
\facilities{Chandra
}


\software{HEAsoft (v6.20; \citealt{heasarc14}), CIAO (v4.11; \citealt{fruscione06}), mkacispback (v2021-07-15; \citealt{suzuki21b})
}

\vspace{30mm}

\appendix
\section{Flux profiles across the filaments}\label{sec-prop2}
\begin{figure*}[htb!]
\centering
\includegraphics[width=14cm]{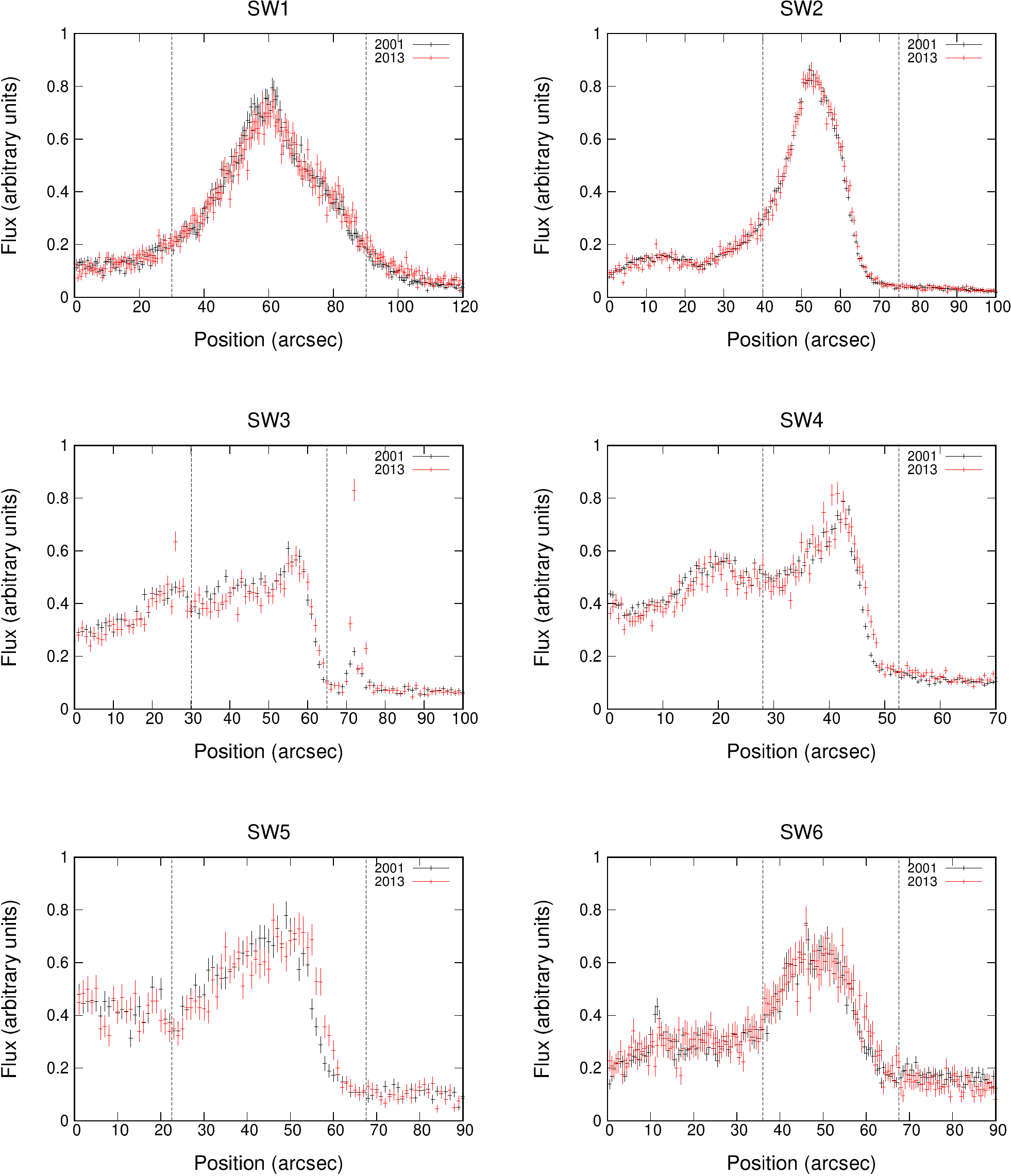}
\caption{Flux profiles of the regions SW1--SW6.
The black and red crosses represent the data taken in 2001and 2013, respectively.
The positive directions of positions correspond to the directions of the arrows shown in Figure~\ref{fig-image}.
{The displayed flux ranges are different for different panels.}
The vertical dashed lines represent the ranges used for proper motion measurement.
\label{fig-profile1}}
\end{figure*}

\begin{figure*}[htb!]
\centering
\includegraphics[width=14cm]{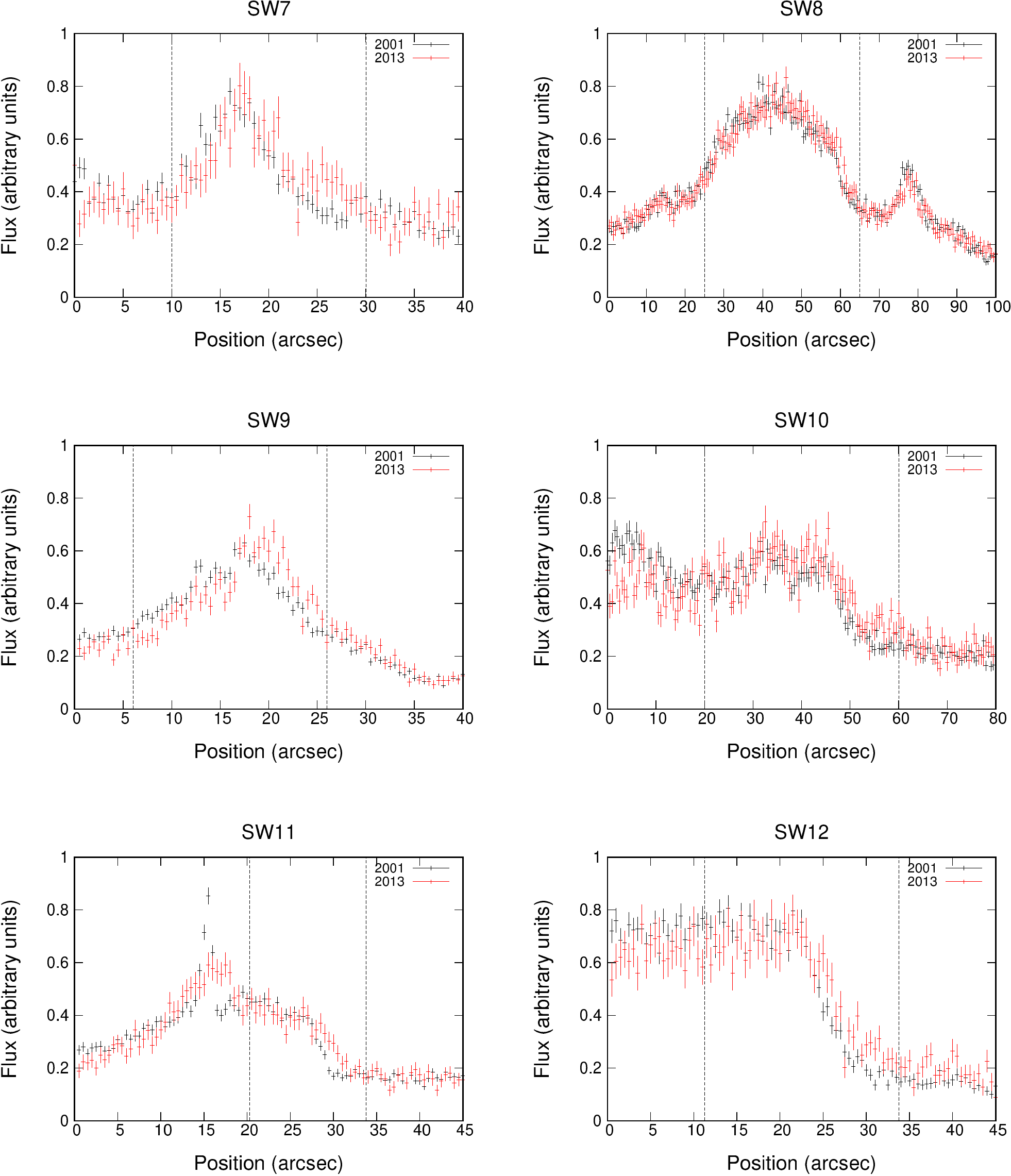}
\caption{Flux profiles of the regions SW7--SW12.
Same convention for the plots is used as in Figure~\ref{fig-profile1}.
\label{fig-profile2}}
\end{figure*}

The X-ray flux profiles of individual filaments in 2001 and 2013 are shown in Figures~\ref{fig-profile1} and \ref{fig-profile2}.
The radial ranges indicated with the dotted lines are used to calculate the proper motion velocities.

\section{Other parameter correlations}\label{sec-other}
\begin{figure*}[htb!]
\centering
\includegraphics[width=16cm]{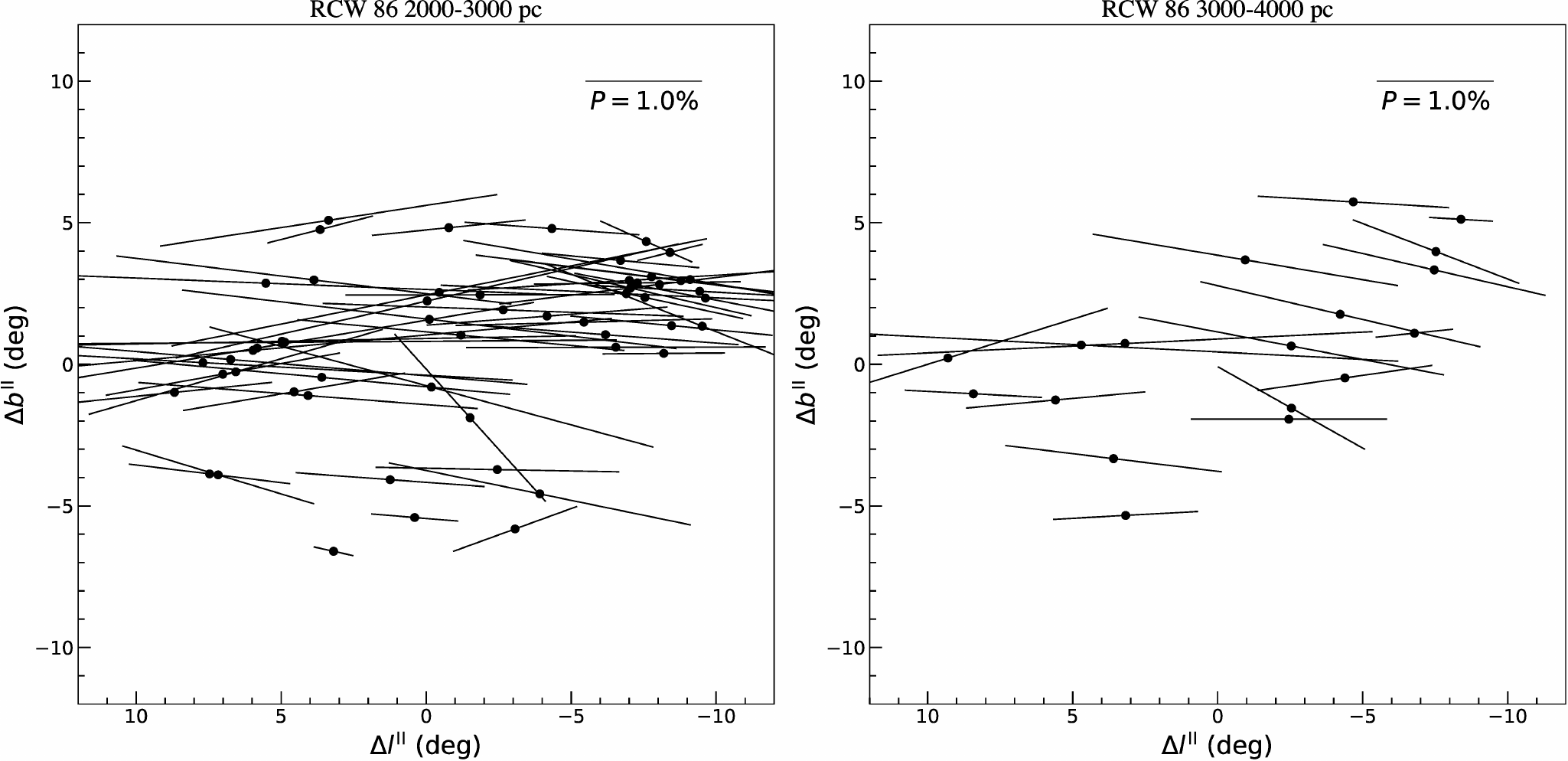}
\caption{Magnetic field directions estimated at the stars around RCW~86 region.
$\Delta l^\parallel$, $\Delta b^\parallel$, and $P$ indicate the Galactic longitude and latitude with respect to the coordinates $(l, b) = (315\fdg4015, -2\fdg31664)$, and polarization degree, respectively.
Left and right panels show the estimates for distances of 2--3~kpc and 3--4~kpc, respectively.
\label{fig-bfield}}
\end{figure*}

\begin{figure*}[htb!]
\centering
\includegraphics[width=16cm]{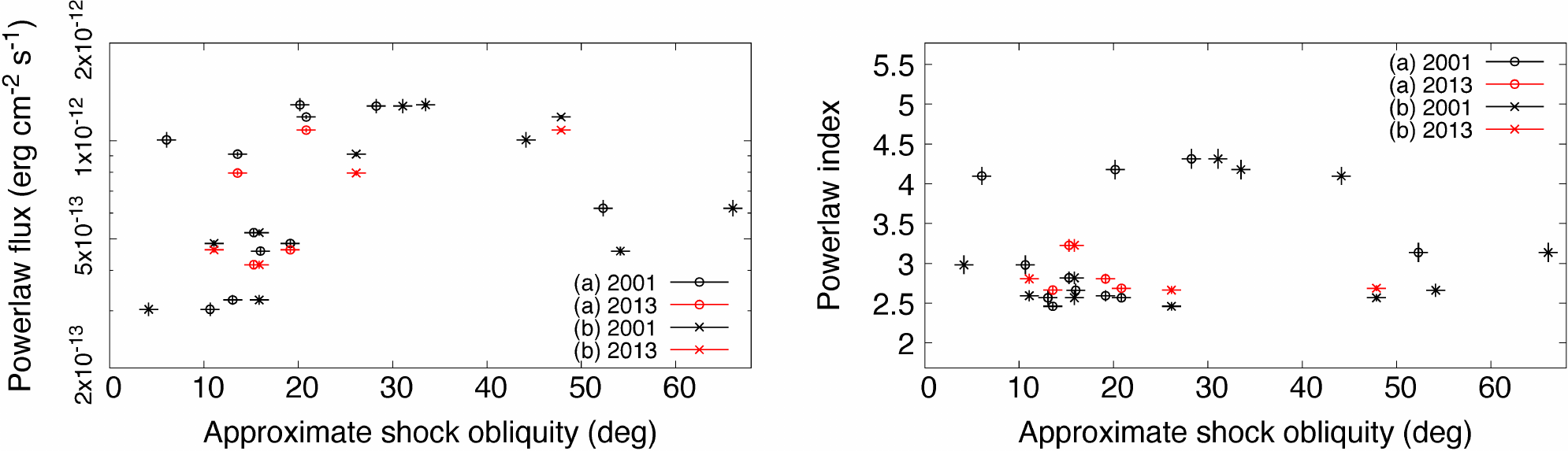}
\caption{Plots of the power-law flux and index against the shock obliquity.
{As a proxy of the shock obliquity, we assume two cases: we use the angle made by the Galactic plane and vector connecting the filament position and geometric center of RCW~86 (case (a)), and the angle made by the Galactic plane and moving direction of the filament defined in Figure~\ref{fig-image} (case (b)).}
For the shock obliquity, a typical error of 1~degree is assigned to all the regions.
\label{fig-pl-obliquity}}
\end{figure*}

In order to investigate the dependence of the non-thermal parameters on the shock obliquity, we first estimate the magnetic field directions as follows.
We use the starlight polarization data compiled by \cite{heiles00}.
This database includes the polarization properties of $\sim 10000$ stars with their positions and distances.
We extract the data of the stars in a 20$^\circ$ $\times$ 20$^\circ$ square region centered on $(l, b) = (315\fdg4015, -2\fdg31664)$ at distances of 2--4~kpc.
The resultant magnetic field directions around RCW~86, which directly correspond to the starlight polarization directions, are presented in Figure~\ref{fig-bfield}.
The directions are found to be nearly parallel to the Galactic plane.
We then calculate the shock obliquity of each filament assuming that the filaments have been moving straight against the explosion center estimated in Section~\ref{sec-inward} (case (a)).
The plots of the power-law parameters over the shock obliquity are presented in Figure~\ref{fig-pl-obliquity}.
No clear correlations are found.
Instead of assuming the simple filament motion against the explosion center, we have also assumed that the shock normal directly corresponds to the moving direction we defined in Figure~\ref{fig-image} (case (b)).
This assumption also results in similar plots (Figure~\ref{fig-pl-obliquity}), without significant correlations.

\bibliography{/Users/suzuki/Documents/paper_submission/references/references.bib}

\begin{thebibliography}{}
\expandafter\ifx\csname natexlab\endcsname\relax\def\natexlab#1{#1}\fi
\providecommand{\url}[1]{\href{#1}{#1}}
\providecommand{\dodoi}[1]{doi:~\href{http://doi.org/#1}{\nolinkurl{#1}}}
\providecommand{\doeprint}[1]{\href{http://ascl.net/#1}{\nolinkurl{http://ascl.net/#1}}}
\providecommand{\doarXiv}[1]{\href{https://arxiv.org/abs/#1}{\nolinkurl{https://arxiv.org/abs/#1}}}

\bibitem[{{Allen} {et~al.}(2004){Allen}, {Jerius}, \& {Gaetz}}]{allen04}
{Allen}, C., {Jerius}, D.~H., \& {Gaetz}, T.~J. 2004, in Society of
  Photo-Optical Instrumentation Engineers (SPIE) Conference Series, Vol. 5165,
  X-Ray and Gamma-Ray Instrumentation for Astronomy XIII, ed. K.~A. {Flanagan}
  \& O.~H.~W. {Siegmund}, 423--432, \dodoi{10.1117/12.509411}

\bibitem[{{Amato} \& {Blasi}(2006)}]{amato06}
{Amato}, E., \& {Blasi}, P. 2006, \mnras, 371, 1251,
  \dodoi{10.1111/j.1365-2966.2006.10739.x}

\bibitem[{{Arnaud}(1996)}]{arnaud96}
{Arnaud}, K.~A. 1996, in Astronomical Society of the Pacific Conference Series,
  Vol. 101, Astronomical Data Analysis Software and Systems V, ed. G.~H.
  {Jacoby} \& J.~{Barnes}, 17

\bibitem[{{Bamba} {et~al.}(2000){Bamba}, {Koyama}, \& {Tomida}}]{bamba00b}
{Bamba}, A., {Koyama}, K., \& {Tomida}, H. 2000, \pasj, 52, 1157,
  \dodoi{10.1093/pasj/52.6.1157}

\bibitem[{{Bamba} {et~al.}(2005){Bamba}, {Yamazaki}, {Yoshida}, {Terasawa}, \&
  {Koyama}}]{bamba05}
{Bamba}, A., {Yamazaki}, R., {Yoshida}, T., {Terasawa}, T., \& {Koyama}, K.
  2005, \apj, 621, 793, \dodoi{10.1086/427620}

\bibitem[{{Bell}(2004)}]{bell04}
{Bell}, A.~R. 2004, \mnras, 353, 550, \dodoi{10.1111/j.1365-2966.2004.08097.x}

\bibitem[{{Dickel} {et~al.}(2001){Dickel}, {Strom}, \& {Milne}}]{dickel01}
{Dickel}, J.~R., {Strom}, R.~G., \& {Milne}, D.~K. 2001, \apj, 546, 447,
  \dodoi{10.1086/318259}

\bibitem[{{Fraschetti}(2013)}]{fraschetti13}
{Fraschetti}, F. 2013, \apj, 770, 84, \dodoi{10.1088/0004-637X/770/2/84}

\bibitem[{{Fraschetti} {et~al.}(2018){Fraschetti}, {Katsuda}, {Sato},
  {Jokipii}, \& {Giacalone}}]{fraschetti18}
{Fraschetti}, F., {Katsuda}, S., {Sato}, T., {Jokipii}, J.~R., \& {Giacalone},
  J. 2018, \prl, 120, 251101, \dodoi{10.1103/PhysRevLett.120.251101}

\bibitem[{Fruscione {et~al.}(2006)Fruscione, McDowell, Allen, Brickhouse,
  Burke, Davis, Durham, Elvis, Galle, Harris, Huenemoerder, Houck, Ishibashi,
  Karovska, Nicastro, Noble, Nowak, Primini, Siemiginowska, Smith, \&
  Wise}]{fruscione06}
Fruscione, A., McDowell, J.~C., Allen, G.~E., {et~al.} 2006, in Observatory
  Operations: Strategies, Processes, and Systems, ed. D.~R. Silva \& R.~E.
  Doxsey, Vol. 6270, International Society for Optics and Photonics (SPIE), 586
  -- 597, \dodoi{10.1117/12.671760}

\bibitem[{{Garmire}(1997)}]{garmire97}
{Garmire}, G.~P. 1997, in American Astronomical Society Meeting Abstracts, Vol.
  190, American Astronomical Society Meeting Abstracts \#190, 34.04

\bibitem[{{Giacalone} \& {Jokipii}(2007)}]{giacalone07}
{Giacalone}, J., \& {Jokipii}, J.~R. 2007, \apjl, 663, L41,
  \dodoi{10.1086/519994}

\bibitem[{{Green} \& {Stephenson}(2003)}]{green03}
{Green}, D.~A., \& {Stephenson}, F.~R. 2003, {Historical Supernovae}, ed.
  K.~{Weiler}, Vol. 598, 7--19, \dodoi{10.1007/3-540-45863-8_2}

\bibitem[{{HEASARC}(2014)}]{heasarc14}
{HEASARC}. 2014, {HEAsoft: Unified Release of FTOOLS and XANADU}.
\newblock \doeprint{1408.004}

\bibitem[{{Heiles}(2000)}]{heiles00}
{Heiles}, C. 2000, \aj, 119, 923, \dodoi{10.1086/301236}

\bibitem[{{Helder} {et~al.}(2012){Helder}, {Vink}, {Bykov}, {Ohira}, {Raymond},
  \& {Terrier}}]{helder12}
{Helder}, E.~A., {Vink}, J., {Bykov}, A.~M., {et~al.} 2012, \ssr, 173, 369,
  \dodoi{10.1007/s11214-012-9919-8}

\bibitem[{{HI4PI Collaboration} {et~al.}(2016){HI4PI Collaboration}, {Ben
  Bekhti}, {Fl{\"o}er}, {Keller}, {Kerp}, {Lenz}, {Winkel}, {Bailin},
  {Calabretta}, {Dedes}, {Ford}, {Gibson}, {Haud}, {Janowiecki}, {Kalberla},
  {Lockman}, {McClure-Griffiths}, {Murphy}, {Nakanishi}, {Pisano}, \&
  {Staveley-Smith}}]{hi4pi16}
{HI4PI Collaboration}, {Ben Bekhti}, N., {Fl{\"o}er}, L., {et~al.} 2016, \aap,
  594, A116, \dodoi{10.1051/0004-6361/201629178}

\bibitem[{{Hickox} \& {Markevitch}(2006)}]{hickox06}
{Hickox}, R.~C., \& {Markevitch}, M. 2006, \apj, 645, 95,
  \dodoi{10.1086/504070}

\bibitem[{{Inoue} {et~al.}(2012){Inoue}, {Yamazaki}, {Inutsuka}, \&
  {Fukui}}]{inoue12}
{Inoue}, T., {Yamazaki}, R., {Inutsuka}, S.-i., \& {Fukui}, Y. 2012, \apj, 744,
  71, \dodoi{10.1088/0004-637X/744/1/71}

\bibitem[{{Katsuda} {et~al.}(2010){Katsuda}, {Petre}, {Mori}, {Reynolds},
  {Long}, {Winkler}, \& {Tsunemi}}]{katsuda10}
{Katsuda}, S., {Petre}, R., {Mori}, K., {et~al.} 2010, \apj, 723, 383,
  \dodoi{10.1088/0004-637X/723/1/383}

\bibitem[{{Kishishita} {et~al.}(2013){Kishishita}, {Hiraga}, \&
  {Uchiyama}}]{kishishita13}
{Kishishita}, T., {Hiraga}, J., \& {Uchiyama}, Y. 2013, \aap, 551, A132,
  \dodoi{10.1051/0004-6361/201220525}

\bibitem[{{Kuntz} \& {Snowden}(2000)}]{kuntz00}
{Kuntz}, K.~D., \& {Snowden}, S.~L. 2000, \apj, 543, 195,
  \dodoi{10.1086/317071}

\bibitem[{{Kushino} {et~al.}(2002){Kushino}, {Ishisaki}, {Morita}, {Yamasaki},
  {Ishida}, {Ohashi}, \& {Ueda}}]{kushino02}
{Kushino}, A., {Ishisaki}, Y., {Morita}, U., {et~al.} 2002, \pasj, 54, 327,
  \dodoi{10.1093/pasj/54.3.327}

\bibitem[{{Masui} {et~al.}(2009){Masui}, {Mitsuda}, {Yamasaki}, {Takei},
  {Kimura}, {Yoshino}, \& {McCammon}}]{masui09}
{Masui}, K., {Mitsuda}, K., {Yamasaki}, N.~Y., {et~al.} 2009, \pasj, 61, S115,
  \dodoi{10.1093/pasj/61.sp1.S115}

\bibitem[{{Pohl} {et~al.}(2005){Pohl}, {Yan}, \& {Lazarian}}]{pohl05}
{Pohl}, M., {Yan}, H., \& {Lazarian}, A. 2005, \apjl, 626, L101,
  \dodoi{10.1086/431902}

\bibitem[{{Ressler} {et~al.}(2014){Ressler}, {Katsuda}, {Reynolds}, {Long},
  {Petre}, {Williams}, \& {Winkler}}]{ressler14}
{Ressler}, S.~M., {Katsuda}, S., {Reynolds}, S.~P., {et~al.} 2014, \apj, 790,
  85, \dodoi{10.1088/0004-637X/790/2/85}

\bibitem[{{Reynolds}(2008)}]{reynolds08}
{Reynolds}, S.~P. 2008, \araa, 46, 89,
  \dodoi{10.1146/annurev.astro.46.060407.145237}

\bibitem[{{Reynolds} {et~al.}(2021){Reynolds}, {Williams}, {Borkowski}, \&
  {Long}}]{reynolds21}
{Reynolds}, S.~P., {Williams}, B.~J., {Borkowski}, K.~J., \& {Long}, K.~S.
  2021, \apj, 917, 55, \dodoi{10.3847/1538-4357/ac0ced}

\bibitem[{{Rho} {et~al.}(2002){Rho}, {Dyer}, {Borkowski}, \&
  {Reynolds}}]{rho02}
{Rho}, J., {Dyer}, K.~K., {Borkowski}, K.~J., \& {Reynolds}, S.~P. 2002, \apj,
  581, 1116, \dodoi{10.1086/344248}

\bibitem[{{Rosado} {et~al.}(1996){Rosado}, {Ambrocio-Cruz}, {Le Coarer}, \&
  {Marcelin}}]{rosado96}
{Rosado}, M., {Ambrocio-Cruz}, P., {Le Coarer}, E., \& {Marcelin}, M. 1996,
  \aap, 315, 243

\bibitem[{{Sano} {et~al.}(2013){Sano}, {Tanaka}, {Torii}, {Fukuda}, {Yoshiike},
  {Sato}, {Horachi}, {Kuwahara}, {Hayakawa}, {Matsumoto}, {Inoue}, {Yamazaki},
  {Inutsuka}, {Kawamura}, {Tachihara}, {Yamamoto}, {Okuda}, {Mizuno}, {Onishi},
  {Mizuno}, \& {Fukui}}]{sano13}
{Sano}, H., {Tanaka}, T., {Torii}, K., {et~al.} 2013, \apj, 778, 59,
  \dodoi{10.1088/0004-637X/778/1/59}

\bibitem[{{Sano} {et~al.}(2015){Sano}, {Fukuda}, {Yoshiike}, {Sato}, {Horachi},
  {Kuwahara}, {Torii}, {Hayakawa}, {Tanaka}, {Matsumoto}, {Inoue}, {Yamazaki},
  {Inutsuka}, {Kawamura}, {Yamamoto}, {Okuda}, {Tachihara}, {Mizuno}, {Onishi},
  {Mizuno}, {Acero}, \& {Fukui}}]{sano15}
{Sano}, H., {Fukuda}, T., {Yoshiike}, S., {et~al.} 2015, \apj, 799, 175,
  \dodoi{10.1088/0004-637X/799/2/175}

\bibitem[{{Sano} {et~al.}(2017){Sano}, {Reynoso}, {Mitsuishi}, {Nakamura},
  {Furukawa}, {Mruganka}, {Fukuda}, {Yoshiike}, {Nishimura}, {Ohama}, {Torii},
  {Kuwahara}, {Okuda}, {Yamamoto}, {Tachihara}, \& {Fukui}}]{sano17}
{Sano}, H., {Reynoso}, E.~M., {Mitsuishi}, I., {et~al.} 2017, Journal of High
  Energy Astrophysics, 15, 1, \dodoi{10.1016/j.jheap.2017.04.002}

\bibitem[{{Sano} {et~al.}(2019){Sano}, {Rowell}, {Reynoso}, {Jung-Richardt},
  {Yamane}, {Nagaya}, {Yoshiike}, {Hayashi}, {Torii}, {Maxted}, {Mitsuishi},
  {Inoue}, {Inutsuka}, {Yamamoto}, {Tachihara}, \& {Fukui}}]{sano19b}
{Sano}, H., {Rowell}, G., {Reynoso}, E.~M., {et~al.} 2019, \apj, 876, 37,
  \dodoi{10.3847/1538-4357/ab108f}

\bibitem[{{Sato} {et~al.}(2018){Sato}, {Katsuda}, {Morii}, {Bamba}, {Hughes},
  {Maeda}, {Ishida}, \& {Fraschetti}}]{sato18}
{Sato}, T., {Katsuda}, S., {Morii}, M., {et~al.} 2018, \apj, 853, 46,
  \dodoi{10.3847/1538-4357/aaa021}

\bibitem[{{Smith}(1997)}]{smith97}
{Smith}, R.~C. 1997, \aj, 114, 2664, \dodoi{10.1086/118676}

\bibitem[{{Snowden} {et~al.}(1992){Snowden}, {Plucinsky}, {Briel}, {Hasinger},
  \& {Pfeffermann}}]{snowden92}
{Snowden}, S.~L., {Plucinsky}, P.~P., {Briel}, U., {Hasinger}, G., \&
  {Pfeffermann}, E. 1992, \apj, 393, 819, \dodoi{10.1086/171549}

\bibitem[{{Stephenson} \& {Green}(2002)}]{stephenson02}
{Stephenson}, F.~R., \& {Green}, D.~A. 2002, Historical supernovae and their
  remnants, 5

\bibitem[{{Suzuki} {et~al.}(2020){Suzuki}, {Bamba}, {Yamazaki}, \&
  {Ohira}}]{suzuki20b}
{Suzuki}, H., {Bamba}, A., {Yamazaki}, R., \& {Ohira}, Y. 2020, \pasj, 72, 72,
  \dodoi{10.1093/pasj/psaa061}

\bibitem[{{Suzuki} {et~al.}(2022){Suzuki}, {Bamba}, {Yamazaki}, \&
  {Ohira}}]{suzuki22a}
---. 2022, \apj, 924, 45, \dodoi{10.3847/1538-4357/ac33b5}

\bibitem[{{Suzuki} {et~al.}(2021){Suzuki}, {Plucinsky}, {Gaetz}, \&
  {Bamba}}]{suzuki21b}
{Suzuki}, H., {Plucinsky}, P.~P., {Gaetz}, T.~J., \& {Bamba}, A. 2021, \aap,
  655, A116, \dodoi{10.1051/0004-6361/202141458}

\bibitem[{{Tran} {et~al.}(2015){Tran}, {Williams}, {Petre}, {Ressler}, \&
  {Reynolds}}]{tran15}
{Tran}, A., {Williams}, B.~J., {Petre}, R., {Ressler}, S.~M., \& {Reynolds},
  S.~P. 2015, \apj, 812, 101, \dodoi{10.1088/0004-637X/812/2/101}

\bibitem[{{Truelove} \& {McKee}(1999)}]{truelove99}
{Truelove}, J.~K., \& {McKee}, C.~F. 1999, \apjs, 120, 299,
  \dodoi{10.1086/313176}

\bibitem[{{Tsubone} {et~al.}(2017){Tsubone}, {Sawada}, {Bamba}, {Katsuda}, \&
  {Vink}}]{tsubone17}
{Tsubone}, Y., {Sawada}, M., {Bamba}, A., {Katsuda}, S., \& {Vink}, J. 2017,
  \apj, 835, 34, \dodoi{10.3847/1538-4357/835/1/34}

\bibitem[{{Tsuji} {et~al.}(2019){Tsuji}, {Uchiyama}, {Aharonian}, {Berge},
  {Higurashi}, {Krivonos}, \& {Tanaka}}]{tsuji19}
{Tsuji}, N., {Uchiyama}, Y., {Aharonian}, F., {et~al.} 2019, \apj, 877, 96,
  \dodoi{10.3847/1538-4357/ab1b29}

\bibitem[{{Tsuji} {et~al.}(2021){Tsuji}, {Uchiyama}, {Khangulyan}, \&
  {Aharonian}}]{tsuji21}
{Tsuji}, N., {Uchiyama}, Y., {Khangulyan}, D., \& {Aharonian}, F. 2021, \apj,
  907, 117, \dodoi{10.3847/1538-4357/abce65}

\bibitem[{{Uchiyama} {et~al.}(2007){Uchiyama}, {Aharonian}, {Tanaka},
  {Takahashi}, \& {Maeda}}]{uchiyama07}
{Uchiyama}, Y., {Aharonian}, F.~A., {Tanaka}, T., {Takahashi}, T., \& {Maeda},
  Y. 2007, \nat, 449, 576, \dodoi{10.1038/nature06210}

\bibitem[{{Vink}(2006)}]{vink06b}
{Vink}, J. 2006, in ESA Special Publication, Vol. 604, The X-ray Universe 2005,
  ed. A.~{Wilson}, 319.
\newblock \doarXiv{astro-ph/0601131}

\bibitem[{{Vink}(2017)}]{vink17}
{Vink}, J. 2017, in Handbook of Supernovae, ed. A.~W. {Alsabti} \& P.~{Murdin},
  2063, \dodoi{10.1007/978-3-319-21846-5\_92}

\bibitem[{{Vink} {et~al.}(2006){Vink}, {Bleeker}, {van der Heyden}, {Bykov},
  {Bamba}, \& {Yamazaki}}]{vink06}
{Vink}, J., {Bleeker}, J., {van der Heyden}, K., {et~al.} 2006, \apjl, 648,
  L33, \dodoi{10.1086/507628}

\bibitem[{{V{\"o}lk} {et~al.}(2005){V{\"o}lk}, {Berezhko}, \&
  {Ksenofontov}}]{volk05}
{V{\"o}lk}, H.~J., {Berezhko}, E.~G., \& {Ksenofontov}, L.~T. 2005, \aap, 433,
  229, \dodoi{10.1051/0004-6361:20042015}

\bibitem[{{Williams} {et~al.}(2011){Williams}, {Blair}, {Blondin}, {Borkowski},
  {Ghavamian}, {Long}, {Raymond}, {Reynolds}, {Rho}, \& {Winkler}}]{williams11}
{Williams}, B.~J., {Blair}, W.~P., {Blondin}, J.~M., {et~al.} 2011, \apj, 741,
  96, \dodoi{10.1088/0004-637X/741/2/96}

\bibitem[{{Yamaguchi} {et~al.}(2016){Yamaguchi}, {Katsuda}, {Castro},
  {Williams}, {Lopez}, {Slane}, {Smith}, \& {Petre}}]{yamaguchi16}
{Yamaguchi}, H., {Katsuda}, S., {Castro}, D., {et~al.} 2016, \apjl, 820, L3,
  \dodoi{10.3847/2041-8205/820/1/L3}

\bibitem[{{Yamaguchi} {et~al.}(2011){Yamaguchi}, {Koyama}, \&
  {Uchida}}]{yamaguchi11}
{Yamaguchi}, H., {Koyama}, K., \& {Uchida}, H. 2011, \pasj, 63, S837,
  \dodoi{10.1093/pasj/63.sp3.S837}

\bibitem[{{Yamazaki} {et~al.}(2014){Yamazaki}, {Ohira}, {Sawada}, \&
  {Bamba}}]{yamazaki14}
{Yamazaki}, R., {Ohira}, Y., {Sawada}, M., \& {Bamba}, A. 2014, Research in
  Astronomy and Astrophysics, 14, 165, \dodoi{10.1088/1674-4527/14/2/005}

\bibitem[{{Yoshino} {et~al.}(2009){Yoshino}, {Mitsuda}, {Yamasaki}, {Takei},
  {Hagihara}, {Masui}, {Bauer}, {McCammon}, {Fujimoto}, {Wang}, \&
  {Yao}}]{yoshino09}
{Yoshino}, T., {Mitsuda}, K., {Yamasaki}, N.~Y., {et~al.} 2009, \pasj, 61, 805,
  \dodoi{10.1093/pasj/61.4.805}

\bibitem[{{Yuan} {et~al.}(2014){Yuan}, {Huang}, {Liu}, \& {Zhang}}]{yuan14}
{Yuan}, Q., {Huang}, X., {Liu}, S., \& {Zhang}, B. 2014, \apjl, 785, L22,
  \dodoi{10.1088/2041-8205/785/2/L22}

\bibitem[{{Zeng} {et~al.}(2019){Zeng}, {Xin}, \& {Liu}}]{zeng19}
{Zeng}, H., {Xin}, Y., \& {Liu}, S. 2019, \apj, 874, 50,
  \dodoi{10.3847/1538-4357/aaf392}

\bibitem[{{Zirakashvili} \& {Aharonian}(2007)}]{zira07}
{Zirakashvili}, V.~N., \& {Aharonian}, F. 2007, \aap, 465, 695,
  \dodoi{10.1051/0004-6361:20066494}

\end{thebibliography}
\bibliographystyle{aasjournal}



\end{document}